\documentclass{myjournal} 

\usepackage[ruled, linesnumbered, lined, noend]{algorithm2e} 

\SetAlFnt{\small}
\SetAlCapFnt{\small}
\SetAlCapNameFnt{\small}
\SetAlCapHSkip{0pt}

\usepackage{amsmath}
\usepackage{amsthm}
\newtheorem{mylemma}{Lemma}
\newcommand{\citet}[1]{\cite{#1}}

\usepackage{xcolor,hyperref}
\usepackage{graphicx}
\usepackage{subcaption}
\usepackage{xspace}

\newcommand{\myparagraph}[1]{\paragraph*{#1}}

\newcommand{\Id}[1]{\textit{#1}}
\newcommand{\Oh}[1]{\mathrm{O}\!\left( #1\right)}
\newcommand{\Om}[1]{\mathrm{\Omega}\!\left(#1\right)}
\newcommand{\set}[1]{\left\{ #1\right\}}
\newcommand{\Is}{\ensuremath{\mathbin{:=}}}
\newcommand{\floor}[1]{\left\lfloor #1\right\rfloor}

\newcommand{\boundedName}{\textit{folklore}}

\definecolor{kit_green}{rgb}{0    , 0.588, 0.509}
\definecolor{kit_blue} {rgb}{0.274, 0.392, 0.666}
\definecolor{kit_red}  {rgb}{0.627, 0.117, 0.156}

\definecolor{hone}{HTML}{222222}
\definecolor{htwo}{HTML}{aaaaaa}

\usepackage{amssymb}
\usepackage{tikz}

\newcommand{\fsqu}{%
\begin{tikzpicture}%
   \filldraw (0,0) rectangle (1.4ex,1.4ex);%
\end{tikzpicture}%
}
\newcommand{\lsqu}{%
\begin{tikzpicture}%
   \draw (0,0) rectangle (1.4ex,1.4ex);%
   \clip (0,0) rectangle (1.4ex,1.4ex);%
   \fill (-0.7ex,-0.7ex) rectangle (0.7ex,1.4ex);%
\end{tikzpicture}%
}
\newcommand{\bsqu}{%
\begin{tikzpicture}%
   \draw (0,0) rectangle (1.4ex,1.4ex);%
   \clip (0,0) rectangle (1.4ex,1.4ex);%
   \fill (-0.7ex,-0.7ex) rectangle (1.4ex,0.7ex);%
\end{tikzpicture}%
}

\newcommand{\fdiam}{%
\begin{tikzpicture}%
   \filldraw (0,0) [rotate=45] rectangle (1.1ex,1.1ex);%
\end{tikzpicture}%
}

\newcommand{\fcirc}{%
\begin{tikzpicture}%
   \filldraw (0,0) circle (0.7ex);%
\end{tikzpicture}%
}
\newcommand{\lcirc}{%
\begin{tikzpicture}%
   \draw (0,0) circle (0.7ex);%
   \clip (0,0) circle (0.7ex);%
   \fill (0ex,0.7ex) rectangle (-0.7ex,-0.7ex);%
\end{tikzpicture}%
}
\newcommand{\bcirc}{%
\begin{tikzpicture}%
   \draw (0,0) circle (0.7ex);%
   \clip (0,0) circle (0.7ex);%
   \fill (0.7ex,0ex) rectangle (-0.7ex,-0.7ex);%
\end{tikzpicture}%
}

\newcommand{\Yup}{%
\begin{tikzpicture}%
  \draw (0,0) -- (0ex,0.8ex);%
  \draw [rotate=120] (0,0) -- (0ex,0.8ex);%
  \draw [rotate=240] (0,0) -- (0ex,0.8ex);%
\end{tikzpicture}%
}

\newcommand{\ck}{$\checkmark$}
\newcommand{\uag}  {{\color{hone} \fsqu}\xspace}
\newcommand{\usg}  {{\color{htwo} \fsqu}\xspace}
\newcommand{\pag}  {{\color{hone} \lsqu}\xspace}
\newcommand{\psg}  {{\color{htwo} \bsqu}\xspace}
\newcommand{\folk} {{\color{hone} \fdiam}\xspace}
\newcommand{\junca}{{\color{hone} \fcirc}\xspace}
\newcommand{\juncb}{{\color{htwo} \bcirc}\xspace}
\newcommand{\juncc}{{\color{htwo} \lcirc}\xspace}
\newcommand{\tbba} {{\color{hone} $\bigstar$}\xspace}
\newcommand{\tbbb} {{\color{htwo} $\bigstar$}\xspace}
\newcommand{\folly}{{\color{hone} $+$}\xspace}
\newcommand{\cuck} {{\color{hone} \Yup}\xspace}
\newcommand{\rcua} {{\color{hone} $\times$}\xspace}
\newcommand{\rcub} {{\color{htwo} $\times$}\xspace}
\newcommand{\phase}{{\color{hone} $\blacklozenge$}\xspace}
\newcommand{\hops} {{\color{hone} $\blacktriangle$}\xspace}
\newcommand{\lea}  {{\color{hone} $\blacktriangledown$}\xspace}

\begin{document}


\title{Concurrent Hash Tables: Fast and General(?)!}

\author{Tobias Maier$^1$ \and Peter Sanders$^1$ \and Roman Dementiev$^2$}
\institute{$^1$ Karlsruhe Institute of Technology, Karlsruhe, Germany \quad \email{\{t.maier,sanders\}@kit.edu}\\
           $^2$ Intel Deutschland GmbH \hspace{4.53cm} \email{roman.dementiev@intel.com}}


\maketitle

\begin{abstract}
  Concurrent hash tables are one of the most important concurrent data
  structures which is used in numerous applications. Since hash table accesses
  can dominate the execution time of whole applications, we need implementations
  that achieve good speedup even in these cases. Unfortunately, currently
  available concurrent hashing libraries turn out to be far away from this
  requirement in particular when adaptively sized tables are necessary or
  contention on some elements occurs.

  Our starting point for better performing data structures is a fast and simple
  lock-free concurrent hash table based on linear probing that is however
  limited to word sized key-value types and does not support dynamic size
  adaptation. We explain how to lift these limitations in a provably scalable
  way and demonstrate that dynamic growing has a performance overhead comparable
  to the same generalization in sequential hash tables.

  We perform extensive experiments comparing the performance of our
  implementations with six of the most widely used concurrent hash tables. Ours
  are considerably faster than the best algorithms with similar restrictions and
  an order of magnitude faster than the best more general tables. In some
  extreme cases, the difference even approaches four orders of magnitude.
\end{abstract}

{\bf Category: }
[D.1.3] Programming Techniques Concurrent Programming
[E.1] Data Structures Tables
[E.2] Data Storage Representation Hash-table representations

{\bf Terms: }
Performance, Experimentation, Measurement, Design, Algorithms


\section{Introduction}\label{sec:int}

A hash table is a dynamic data structure which stores a set of elements that are
accessible by their key. It supports insertion, deletion, find and update in
constant expected time. In a concurrent hash table, multiple threads have access
to the same table.  This allows threads to share information in a flexible and
efficient way. Therefore, concurrent hash tables are one of the most important
concurrent data structures. See Section~\ref{sec:semantics} for a more detailed
discussion of concurrent hash table functionality.

To show the ubiquity of hash tables we give a short list of example
applications: A very simple use case is storing sparse sets of precomputed
solutions (e.g. \cite{precomPassHash}, 
\cite{BFMSS07}).  A more complicated one
is aggregation as it is frequently used in analytical data base queries of the
form {\tt SELECT FROM}\ldots {\tt COUNT}\ldots {\tt GROUP BY} $x$
\cite{muller2015cache}.  Such a query selects rows from one or several relations
and counts for every key $x$ how many rows have been found (similar queries work
with {\tt SUM}, {\tt MIN}, or {\tt MAX}).  Hashing can also be used for a
data-base join \cite{chen2007improving}.  Another group of examples is the
exploration of a large combinatorial search space where a hash table is used to
remember the already explored elements (e.g., in dynamic programming
\cite{StivalaCASTable10}, itemset mining \cite{PCY95}, a chess program, or when
exploring an implicitly defined graph in model checking
\cite{stornetta1996implementation}). Similarly, a hash table can maintain a set
of cached objects to save I/Os \cite{nishtala2013scaling}.  Further examples are
duplicate removal, storing the edge set of a sparse graph in order to support
edge queries \cite{MehSan08}, maintaining the set of nonempty cells in a
grid-data structure used in geometry processing
(e.g. \cite{dietzfelbinger1997reliable}), or maintaining the children in tree
data structures such as van Emde-Boas search trees \cite{DKMS04} or suffix trees
\cite{McCreight:1976:SES}.

Many of these applications have in common that -- even in the sequential version
of the program -- hash table accesses constitute a significant fraction of the
running time. Thus, it is essential to have highly scalable concurrent hash
tables that actually deliver significant speedups in order to parallelize these
applications.  Unfortunately, currently available general purpose concurrent
hash tables do not offer the needed scalability (see Section~\ref{sec:exp} for
concrete numbers). On the other hand, it seems to be folklore that a lock-free
linear probing hash table where keys and values are machine words, which is
preallocated to a bounded size, and which supports no true deletion operation
can be implemented using atomic compare-and-swap (CAS) instructions
\cite{StivalaCASTable10}.  Find-operations can even proceed naively and without
any write operations.  In Section~\ref{sec:nongrow} we explain our own
implementation (\boundedName) in detail, after elaborating on some related work,
and introducing the necessary notation (in \autoref{sec:rel} and
\ref{sec:prelim} respectively).
To see the potential big performance differences, consider an exemplary
situation with mostly read only access to the table and heavy contention for a
small number of elements that are accessed again and again by all threads.
\boundedName\ actually profits from this situation because the contended
elements are likely to be replicated into local caches.  On the other hand, any
implementation that needs locks or CAS instructions for find-operations, will
become much slower than the sequential code on current machines.  The purpose of
our paper is to document and explain performance differences, and, more
importantly, to explore to what extent we can make \boundedName\ more general
with an acceptable deterioration in performance.

These generalizations are discussed in \autoref{sec:general}.  We explain how to
grow (and shrink) such a table, and how to support deletions and more general
data types.  In \autoref{sec:add_tsx} we explain how hardware transactional
memory can be used to speed up insertions and updates and how it may help to
handle more general data types.

After describing implementation details in \autoref{sec:impl}, \autoref{sec:exp}
experimentally compares our hash tables with six of the most widely used
concurrent hash tables for microbenchmarks including insertion, finding, and
aggregating data.  We look at both uniformly distributed and skewed input
distributions.  \autoref{sec:conclusion} summarizes the results and discusses
possible lines of future research.

\section{Related Work}\label{sec:rel}
This publication follows up on our previous findings about generalizing fast
concurrent hash tables~\cite{ppoppShort}.  In addition to
describing how to generalize a fast linear probing hash table, we offer an
extensive experimental analysis comparing many concurrent hash tables from
several libraries.

There has been extensive previous work on concurrent hashing.  The widely used
textbook ``The Art of Multiprocessor Programming'' 
\cite{HerSha12} by Herlihy and Shavit devotes an entire chapter to
concurrent hashing and gives an overview over previous work. However, it seems
to us that a lot of previous work focuses more on concepts and correctness but
surprisingly little on  scalability. 
For example, most of the discussed growing mechanisms assume that the
size of the hash table is known exactly without a discussion that this
introduces a performance bottleneck limiting the speedup to a
constant. Similarly, the actual migration is often done sequentially.

Stivala et al.~\cite{StivalaCASTable10} describe a bounded concurrent linear
probing hash table specialized for dynamic programming that only support insert
and find. Their insert operation starts from scratch when the CAS fails which
seems suboptimal in the presence of contention. An interesting point is that
they need only word size CAS instructions at the price of reserving a special
empty value. This technique could also be adapted to port our code to machines
without 128-bit CAS.

Kim and Kim \cite{kim2013performance} compare this table with a cache-optimized
lockless implementation of hashing with chaining and with hopscotch hashing
\cite{herlihy2008hopscotch}. The experiments use only uniformly distributed
keys, i.e., there is little contention.  Both linear probing and hashing with
chaining perform well in that case. The evaluation of find-performance is a bit
inconclusive: chaining wins but using more space than linear probing. Moreover
it is not specified whether this is for successful (use key of inserted
elements) or mostly unsuccessful (generate fresh keys) accesses.  We suspect
that varying these parameters could reverse the result.

Gao et al.~\cite{GaoGrooteHesselink05} present a theoretical dynamic linear
probing hash table, that is lock-free. The main contribution is a formal
correctness proof. Not all details of the algorithm or even an implementation is
given. There is also no analysis of the complexity of the growing procedure.

Shun and Blelloch \cite{shun2014phase} propose \emph{phase concurrent hash tables} which are
allowed to use only a single operation within a globally synchronized
phase. They show how phase concurrency helps to implement some operations more
efficiently and even deterministically in a linear probing context.  For
example, deletions can adapt the approach from \cite{Knu98} and rearrange
elements. This is not possible in a general hash table since this might cause
find-operations to report false negatives.  They also outline an elegant growing
mechanism albeit without implementing it and without filling in all the detail
like how to initialize newly allocated tables.  They propose to trigger a
growing operation when any operation has to scan more than $k\log n$ elements
where $k$ is a tuning parameter. This approach is tempting since it is somewhat
faster than the approximate size estimator we use. We actually tried that but
found that this trigger has a very high variance -- sometimes it triggers late
making operations rather slow, sometimes it triggers early wasting a lot of
space. We also have theoretical concerns since the bound $k\log n$ on the length
of the longest probe sequence implies strong assumptions on certain properties
of the hash function.  Shun and Blelloch make extensive experiments including
applications from the problem based benchmark suite
\cite{shun2012brief}. 

Li et al.~\cite{AlgoImpCuckoo14} use
the bucket cuckoo-hashing method by Dietzfelbinger and Weidling \cite{DieWei07} and develop a concurrent
implementation. They exploit that using a BFS-based insertion algorithm, the
number of element moves for an insertion is very small.  They use fine grained
locks which can sometimes be avoided using transactional memory (Intel
TSX). As a result of their work, they implemented the small open source library libcuckoo,
which we measure against (which does not use TSX). This approach has the
potential to achieve very good space efficiency. However, our measurements
indicate that the performance penalty is high.

The practical importance of concurrent hash tables also leads to new and
innovative implementations outside of the scientific community. A good example
of this is the Junction library, that was published by Preshing \cite{junctionGit}
in the beginning of 2016, shortly after our initial publication \cite{hashTRArxiv}.

\section{Preliminaries}\label{sec:prelim}

We assume that each application thread has its own designated hardware thread or
processing core and denote the number of these threads with $p$.  A data
structure is non-blocking if no blocked thread currently accessing this data
structure can block an operation on the data structure by another thread. A data
structure is lock-free if it is non-blocking and guarantees global progress,
i.e., there must always be at least one thread finishing its operation in a
finite number of steps.

\emph{Hash Tables} store a set of $\langle\Id{Key}, \Id{Value}\rangle$ pairs
(elements).%
\footnote{Much of what is said here can be generalized to the case when
  \Id{Element}s are black boxes from which keys are extracted by an accessor
  function.}
A hash function $h$ maps each key to a cell of a table (an array).  The
number of elements in the hash table is denoted $n$ and the number of operations
is $m$. For the purpose of algorithm analysis, we assume that $n$ and $m$ are
$\gg p^2$ -- this allows us to simplify algorithm complexities by hiding $O(p)$
terms that are independent of $n$ and $m$ in the overall cost. 
Sequential hash tables support the insertion of elements, and finding, updating,
or deleting an element with given key -- all of this in constant expected
time. Further operations compute $n$ (\Id{size}), build a table with a given
number of initial elements, and iterate over all elements (\Id{forall}).

\emph{Linear Probing} is one of the most popular sequential hash table
schemes used in practice.  An element $\langle x,a\rangle$ is stored at the first free table
entry following position $h(x)$ (wrapping around when the end of the table is
reached). Linear probing is at the same time simple and efficient -- if the
table is not too full, a single cache line access will be enough most of the
time. Deletion can be implemented by rearranging the elements locally
\cite{Knu98} to avoid holes violating the invariant mentioned above. When the
table becomes too full or too empty, the elements can be migrated to a larger or
smaller table respectively.  The migration cost can be charged to insertions and
deletions causing amortized constant overhead.

\section{Concurrent Hash Table Interface and Folklore Implementation}
\label{sec:semantics}\label{sec:nongrow}
Although it seems quite clear what a hash table is and how this generalizes to
concurrent hash tables, there is a surprising number of details to
consider. Therefore, we will quickly go over some of our interface decisions,
and detail how this interface can be implemented in a simple, fast, lock-free
concurrent linear probing hash table.

This hash table will have a bounded capacity $c$ that has to be specified when
the table is constructed.  It is the basis for all other hash table variants
presented in this publication. We call this table the \emph{folklore} solution,
because variations of it are used in many publications 
and it is not clear to us by whom it was first published.

The most important requirement for concurrent data structures is, that they
should be \emph{linearizable}, i.e., it must be possible to order the hash table
operations in some sequence -- without reordering two opperations of the same
thread -- so that executing them sequentially in that order yields the same
results as the concurrent processing. For a hash table data structure, this
basically means that all operations should be executed atomically some time
between their invokation and their return. For example, it has to be avoided,
that a \texttt{find} returns an inconsistent state, e.g.~a half-updated data field
that was never actually stored at the corresponding key.

Our variant of the folklore solution ensures the atomicity of operations using
2-word atomic CAS operations for all changes of the table. As long as the key
and the value each only use one machine word, we can use 2-word CAS opearations
to atomically manipulate a stored key together with the corresponding
value. There are other variants that avoid need 2-word compare and swap
operations, but they often need a designated empty value (see \cite{junctionGit})
. Since, the corresponding machine instructions are widely available on modern
hardware, using them should not be a problem. If the target architecture does not
support the needed instructions, the implementation can easily be switched to use
a variant of the folklore solution which does not use 2-word CAS. As it can
easily be deduced by the context, we will usually omit the ``2-word'' prefix and
use the abbreviation CAS for both single and double word CAS operations.




\begin{algorithm*}[t]

\caption{Pseudocode for the \texttt{insertOrUpdate} operation}
\label{alg:mod}

\Indm
\KwIn{Key $k$, Data Element $d$, Update Function $\textit{up}:\ \textit{Key}\times \textit{Val} \times \textit{Val} \rightarrow \textit{Val}$}
\KwOut{Boolean \texttt{true} when a new key was inserted, \texttt{false} if an update occurred}
\Indp
\vspace{0.1cm}
$i$\ \texttt{=}\ \texttt{h(}$k$\texttt{)}\;
\While{\texttt{true}} {
  $i\ \texttt{=}\ i\ \texttt{\%}\ c$\;
  $\textit{current}\ \texttt{=}\ \textit{table}\texttt{[}i\texttt{]}$\;

  \vspace{0.1cm}
  \uIf(\tcp*[f]{\label{l:upretf}Key is not present yet \dots}){\textit{current.key} == \texttt{empty\_key}}{
    \eIf{\textit{table}\texttt{[}$i$\texttt{].CAS(}\textit{current}$, \langle k, d\,\rangle$\texttt{)}} 
        {\Return{\texttt{true}}} 
        { $i$\texttt{-\,\!-}\; }
    \vspace{0.1cm}
  }
  \uElseIf(\tcp*[f]{\label{l:inretf}Same key already present \dots}){$\textit{current.key}\ \texttt{==}\ k$}{
    \eIf{\textit{table}\texttt{[}$i$\texttt{].atomicUpdate(}$\textit{current},\ d,\ \textit{up}$\texttt{)}}
        { 
          \tcp*[f]{\label{l:atomup} default: 
            {\it\texttt{atomicUpdate(}$\cdot$\texttt{) = CAS(}
             $current$\texttt{, }\textit{up}\texttt{(}
             $k, \textit{current.data},\ d$\texttt{))}}}\\
          \Return{false} \label{l:uprett} 
        }
        { $i$\texttt{-\,\!-}\; }
    \vspace{0.1cm}
  }
  \vspace{0.1cm}
  $i$\texttt{++}\;
}
\end{algorithm*}


\paragraph*{Initialization} The constructor allocates an array of size $c$
consisting of 128-bit aligned cells whose key is initialized to the empty
values.

\paragraph*{Modifications} We propose, to categorize all changes to the hash
table content into one of the following three functions, that can be implemented
very similarly (does not cover deletions).

\vspace{0.05cm}
\noindent\texttt{insert(}$k, d$\texttt{)}: Returns \texttt{false} if an element with
the specified key is already present. Only one operation should succeed if
multiple threads are inserting the same key at the same time. 

\vspace{0.05cm}
\noindent\texttt{update(}$k, d, \textit{up}$\texttt{)}: Returns \texttt{false}, if there is no
value stored at the specified key, otherwise this function atomically updates
the stored value to $\textit{new} = \textit{up}(\textit{current}, d)$. Notice,
that the resulting value can be dependent on both the current value and the
input parameter $d$.

\vspace{0.05cm}
\noindent\texttt{insertOrUpdate(}$k, d, \textit{up}$\texttt{)}: This operation updates the
current value, if one is present, otherwise the given data element is inserted
as the new value. The function returns \texttt{true}, if \texttt{insertOrUpdate}
performed an \texttt{insert} (key was not present), and \texttt{false} if an
\texttt{update} was executed.

\vspace{0.05cm} We choose this interface for two main reasons. It allows
applications to quickly differentiate between inserting and changing an element
-- this is especially usefull since the thread who first inserted a key can be
identified uniquely. Additionally it allows transparent, lockless updates that
can be more complex, than just replacing the current value (think of CAS or Fetch-and-Add).

The update interface using an update function deserves some special attention,
as it is a novel approach compared to most interfaces we encountered during our
research. Most implementations fall into one of two categories: They return
mutable references to table elements -- forcing the user to implement atomic
operations on the data type; or they offer an \texttt{update} function which
usually replaces the current value with a new one -- making it very hard to
implement atomic changes like a simple counter (\texttt{find} +
\texttt{increment} + \texttt{overwrite} not necessarily atomic).

In Algorithm~\ref{alg:mod} we show the pseudocode of the \texttt{insertOrUpdate}
function. The operation computes the hash value of the key and proceeds to look for an
element with the appropriate key (beginning at the corresponding position). If
no element matching the key is found (when an empty space is encountered), the
new element has to be inserted. This is done using a CAS operation. A failed
swap can only be caused by another insertion into the same cell. In this case,
we have to revisit the same cell, to check if the inserted element matches the
current key.  If a cell storing the same key is found, it will be updated using
the \texttt{atomicUpdate} function. This function is usually implemented by
evaluating the passed update function (\textit{up}) and using a CAS operation,
to change the cell. In the case of multiple concurrent updates, at least one
will be successful.

In our (C++) implementation, partial template specialization can be used to
implement more efficient \texttt{atomicUpdate} variants using atomic operations
-- changing the default \autoref{l:atomup}, e.g.~overwrite (using single word
store), increment (using fetch and add).

The code presented in Algorithm~\ref{alg:mod} can easily be modified to implement the
\texttt{insert} (return \texttt{false} when the key is already present -- \autoref{l:inretf}) and
\texttt{update} (return \texttt{true} after a successful update -- \autoref{l:uprett} and
\texttt{false} when the key is not found -- \autoref{l:upretf}) functions. All modification functions
have a constant expected running time.

\paragraph*{Lookup} Since this folklore implementation does not move elements
within the table, it would be possible for \texttt{find($k$)} to return a
reference to the corresponding element. In our experience, returning references
directly tempts inexperienced programmers to opperate on these references in a
way that is not necessarily threadsafe. Therefore, our implementation returns a
copy of the corresponding cell ($\langle k, d\,\rangle$), if one is found
($\langle\texttt{empty\_key}, \cdot \rangle$ otherwise). The \texttt{find}
operation has a constant expected running time.

Our implementation of \texttt{find} somewhat non-trivial, because it is not
possible to read two machine words at once using an atomic
instruction\footnote{The element is not read atomically, because x86 does not
  support that. One could use a 2-word CAS to achieve the same effect but this
  would have disastrous effects on performance when many threads try to find the
  same element.}. Therefore it is possible for a cell to be changed inbetween
reading its key and its value -- this is called a \emph{torn read}. We have to
make sure, that torn reads cannot lead to any wrong behavior. There are two
kinds of interesting torn reads: First an empty key is read while the searched
key is inserted into the same cell, in this case the element is not found
(consistent since it has not been fully inserted); Second the element is updated
between the key being read and the data being read, since the data is read
second, only the newer data is read (consistent with a finished update).

\paragraph*{Deletions} The folklore solution can only handle deletions using
dummy elements -- called tombstones. Usually the key stored in a cell is
replaced with \texttt{del\_key}. Afterwards the cell cannot be used anymore.
This method of handling deleted elements is usually not feasible, as it does not
increase the capacity for new elements. In \autoref{ss:delete} We will show, how
our generalizations can be used to handle tombstones more efficiently.

\paragraph*{Bulk Operations} While not often used in practice, the folklore table
can be modified to support operations like \texttt{buildFrom($\cdot$)} (see
\autoref{ss:bulk}) -- using a bulk insertion which can be more efficient than
element-wise insertion -- or \texttt{forall($f$)} -- which can be implemented embarrassingly  parallel by splitting the table between threads.

\paragraph*{Size} Keeping track of the number of contained elements deserves
special notice here because it turns out to be significantly harder in
concurrent hash tables. In sequential hash tables, it is trivial to count the
number of contained elements -- using a single counter. This same method is
possible in parallel tables using atomic fetch and add operations, but it
introduces a massive amount of contention on one single counter creating a
performance bottleneck.

Because of this we did not include a counting method in folklore implementation. In
\autoref{ss:size} we show how this can be alleviated using an approximate count.

\section{Generalizations and Extensions}\label{sec:general}
In this section, we detail how to adapt the concurrent hash table implementation
-- described in the previous section -- to be universally applicable to all hash
table workloads. Most of our efforts have gone into a scalable migration method
that is used to move all elements stored in one table into another table. It
turns out that a fast migration can solve most shortcomings of the folklore
implementation (especially deletions and adaptable size). 

\subsection{Storing Thread-Local Data}\label{ss:handle}
By itself, storing thread specific data connected to a hash table does not offer
additional functionality, but it is necessary to efficiently implement some of
our other extensions. Per-thread data can be used in many different ways, from
counting the number of insertions to caching shared resources.

From a theoretical point of view, it is easy to store thread specific data. The
additional space is usually only dependent on the number of threads ($\Oh{p}$
additional space), since the stored data is often constant sized. Compared to
the hash table this is usually negligible ($p \ll n < c$).

Storing thread specific data is challenging from a software design and performance
perspective. Some of our competitors use a \texttt{register($\cdot$)} function
that each thread has to call before accessing the table.  This allocates some
memory, that can be accessed using the global hash table object.

Our solution uses explicit handles. Each thread has to create a handle, before
accessing the hash table.  These handles can store thread specific data, since
they are not shared between threads.  This is not only in line with the RAII
idiom (resource acquisition is initialization~\cite{meyers2005effective}), it also protects our
implementation from some performance pitfalls like unnecessary indirections and
false sharing\footnote{Significant slow down created by the cache coherency
  protocol due to multiple threads repeatedly changing distinct values within
  the same cache line.}.  Moreover, the data can easily be deleted once the
thread does not use the hash table anymore (delete the handle).



\subsection{Approximating the Size}\label{ss:size}
Keeping an exact count of the elements stored in the hash table can often lead
to contention on one count variable. Therefore, we propose to support only an
approximative size operation.

To keep an approximate count of all elements, each thread maintains a local
counter of its successful insertions (using the method desribed in
\autoref{ss:handle}).  Every $\Theta(p)$ such insertions this counter is
atomically added to a global insertion counter $I$ and then reset. Contention at
$I$ can be provably made small by randomizing the exact number of local
insertions accepted before adding to the global counter, e.g., between $1$ and
$p$. $I$ underestimates the size by at most $\Oh{p^2}$. Since we assume the size
to be $\gg p^2$ this still means a small relative error. By adding the maximal
error, we also get an upper bound for the table size.

If deletions are also allowed, we maintain a global counter $D$ in a similar
way. $S=I-D$ is then a good estimate of the total size as long as $S\gg
p^2$.

When a table is migrated for growing or shrinking (see \autoref{ss:growShrink}), each
migration thread locally counts the elements it moves. At the end of the migration, local
counters are added to create the initial count for $I$ ($D$ is set to $0$).

This method can also be extended to give an exact count -- in absence of
concurrent insertions/deletions. To do this, a list of all handles has to be
stored at the global hash table object. A thread can now iterate over all
handles computing the actual element size.

\subsection{Table Migration}

While Gao et al.~\cite{GaoGrooteHesselink05} have shown that lock-free dynamic
linear probing hash tables are possible, there is no result on their practical
feasibility. Our focus is geared more towards engineering the fastest migration
possible, therefore, we are fine with small amounts of locking, as long as it
improves the overall performance. 

\subsubsection{Eliminating Unnecessary Contention from the Migration}
\label{ss:growShrink}

If the table size is not fixed, it makes sense to assume that the hash function
$h$ yields a large pseudorandom integer which is then mapped to a cell position
in $0..c-1$ where $c$ is the current capacity $c$.\footnote{We use $x..y$ as a
shorthand for $\set{x,\ldots,y}$ in this paper.}  We will discuss a way to do
this by scaling.  If $h$ yields values in the global range $0..U-1$ we map key
$x$ to cell $h_c(x)\Is\lfloor h(x)\frac{c}{U}\rfloor$. Note that when both $c$ and $U$
are powers of two, the mapping can be implemented by a simple shift operation.

\begin{figure*}
\begin{subfigure}[t]{0.48\textwidth}
\centering
\includegraphics[height=3.5cm]{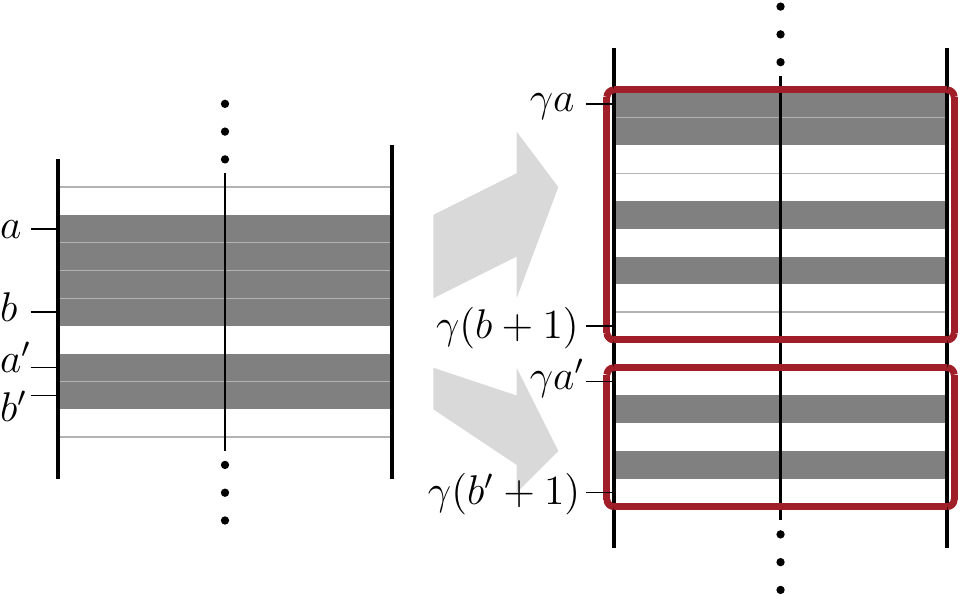}
\caption{Two neighboring clusters and their non-overlapping target areas ($\gamma = 2$).}
\label{fig:cluster}
\end{subfigure}
\quad
\begin{subfigure}[t]{0.48\textwidth}
\centering
\includegraphics[height=3.5cm]{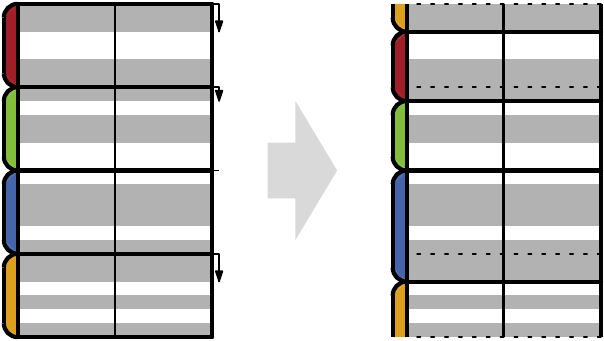}
\caption{Left: table split into even blocks. Right: resulting cluster distribution (moved implicit block borders).}
\label{fig:blocks}
\end{subfigure}
\caption{Cluster migration and work distribution}
\end{figure*}

\myparagraph{Growing} Now suppose that we want to migrate the table into a table
that has at least the same size (growing factor $\gamma\geq 1$).  Exploiting the
properties of linear probing and our scaling function, there is a surprisingly
simple way to migrate the elements from the old table to the new table in
parallel which results in exactly the same order a sequential algorithm would
take and that completely avoids synchronization between threads.
\begin{mylemma}\label{lem:grow} Consider a range $a..b$ of nonempty cells in the
  old table with the property that the cells $a-1\bmod c$ and $b+1\bmod c$ are both
  empty -- call such a range a \emph{cluster} (see \autoref{fig:cluster}). When
  migrating a table, sequential migration will map the elements stored in that
  cluster into the range $\floor{\gamma a}.. \floor{\gamma (b+1)}$
  in the target table, regardless of the rest of the source array.
\end{mylemma} 
\begin{proof}
  Let $x$ be an element stored in the cluster $a..b$ at position
  $p(x) = h_c(x)+d(x)$. Then $h_c(x)$ has to be in the cluster $a..b$, because
  linear probing does not displace elements over empty cells
  ($h_c(x) = \lfloor h(x)\frac{c}{U}\rfloor \geq a$), and therefore,
  $h(x)\frac{c'}{U} \geq a\frac{c'}{c} \geq \gamma a$.
  
  Similarly, from $\lfloor h(x)\frac{c}{U} \rfloor\leq b$ follows
  $h(x)\frac{c}{U} < b+1$, and therefore, $h(x)\frac{c'}{U} < \gamma (b+1)$.
\end{proof}

Therefore, two distinct clusters in the source table cannot overlap in the
target table.  We can exploit this lemma by assigning entire clusters to
migrating threads which can then process each cluster completely independently.
Distributing clusters between threads can easily be achieved by first splitting
the table into blocks (regardless of the tables contents) which we assign to
threads for parallel migration.  A thread assigned block $d..e$ will migrate
those clusters that start within this range -- implicitly moving the block
borders to free cells as seen in \autoref{fig:blocks}).  Since the average
cluster length is short and $c=\Om{p^2}$, it is sufficient to deal out blocks of
size $\Om{p}$ using a single shared global variable and atomic fetch-and-add
operations.  Additionally each thread is responsible for initializing all cells
in its region of the target table.  This is important, because sequentially
initializing the hash table can quickly become infeasible.

Note that waiting for the last thread at the end of the migration introduces
some waiting (locking), but this does not create significant work imbalance,
since the block/cluster migration is really fast and clusters are expected to be
short. 

\myparagraph{Shrinking} Unfortunately, the nice structural Lemma~\ref{lem:grow}
no longer applies. We can still parallelize the migration with little
synchronization. Once more, we cut the source table into blocks that we assign
to threads for migration.  The scaling function maps each block $a..b$ in the
source table to a block $a'..b'$ in the target table.  We have to be careful
with rounding issues so that the blocks in the target table are non-overlapping.
We can then proceed in two phases. First, a migrating thread migrates those
elements that move from $a..b$ to $a'..b'$. These migrations can be done in a
sequential manner, since target blocks are disjoint. The majority of elements
will fit into the target block. Then, after a barrier synchronization, all
elements that did not fit into their respective target blocks are migrated using
concurrent insertion i.e., using atomic operations. This has negligible overhead
since elements like this only exist at the boundaries of blocks.  The resulting
allocation of elements in the target table will no longer be the same as for a
sequential migration but as long as the data structure invariants of a linear
probing hash table are fulfilled, this is not a problem.

\subsubsection{Hiding the Migration from the Underlying Application}
\label{ss:async}

To make the concurrent hash table more general and easy to use, we would like to
avoid all explicit synchronization.  The growing (and shrinking) operations
should be performed asynchronously when needed, without involvement of the
underlying application.
The migration is triggered once the table is filled to a factor $\geq\alpha$
(e.g. $50\,\%$), this is estimated using the approximate count from
\autoref{ss:size}, and checked whenever the global count is updated.  When a
growing operation is triggered, the capacity will be increased by a factor of
$\gamma\geq1$ (Usually $\gamma = 2$).  The difficulty is ensuring that this
operation is done in a transparent way without introducing any inconsistent
behavior and without incurring undue overheads.

To hide the migration process from the user, two problems have to be solved.
First, we have to find threads to grow the table, and second, we have to ensure,
that changing elements in the source table will not lead to any inconsistent
states in the target table (possibly reverting changes made during the
migration).  Each of these problems can be solved in multiple ways.  We
implemented two strategies for each of them resulting in four different variants
of the hash table (mix and match).

\myparagraph{Recruiting User-Threads} A simple approach to dynamically
allocate threads to growing the table, is to ``enslave'' threads that
try to perform table accesses that would otherwise have to wait for the
completion of the growing process anyway.  This works really well when the table
is regularly accessed by all user-threads, but is inefficient in the worst case when
most threads stop accessing the table at some point, e.g., waiting for the
completion of a global computation phase at a barrier. The few threads still
accessing the table at this point will need a lot of time for growing (up to
$\Om{n}$) while most threads are waiting for them. One could try to also enslave
waiting threads but it looks difficult to do this in a sufficiently general and
portable way.

\myparagraph{Using a Dedicated Thread Pool} A provably efficient approach is to
maintain a pool of $p$ threads dedicated to growing the table. They are blocked
until a growing operation is triggered.  This is when they are awoken to
collectively perform the migration in time $\Oh{n/p}$ and then get back to
sleep. During a migration, application threads might have to sleep until the
migration threads are finished. This will increase the CPU time of our migration
threads making this method nearly as efficient as the enslavement variant.
Using a reasonable computation model, one can show that using thread pools for
migration increases the cost of each table access by at most a constant in a
globally amortized sense (over the non-growing folklore solution). We omit the
relatively simple proof.

To remain fair to all competitors, we used exactly as many threads for the
thread pool as there were application threads accessing the table. Additionally
each migration thread was bound to a core, that was also used by one
corresponding application thread.

\myparagraph{Marking Moved Elements for Consistency (asynchronous)}  During the
migration it is important that no element can be changed in the old table after
it has been copied to the new table. Otherwise, it would be hard to
guarantee that changes are correctly applied to the new table. The easiest
solution to this problem is, to mark each cell before it is copied. Marking each
cell can be done using a CAS operation to set a special marked bit which is
stored in the key. In practice this reduces the possible key space. If this
reduction is a problem, see \autoref{ss:restoring_key_space} on how to circumvent
it. To ensure that no copied cell can be changed, it suffices to ensure that no
marked cell can be changed. This can easily be done by checking the bit before
each writing operation, and by using CAS operations for each update. This
prohibits the use of fast atomic operations to change element values.

After the migration, the old hash table has to be deallocated. Before
deallocating an old table, we have to make sure that no thread is currently
using it anymore. This problem can generally be solved by using reference
counting. Instead of storing the table with a usual pointer, we use a reference
counted pointer (e.g.~\texttt{std::shared\_ptr}) to ensure that the table is
eventually freed.

The main disadvantage of counting pointers is that acquiring a counting pointer
requires an atomic increment on a shared counter. Therefore, it is not feasible
to acquire a counting pointer for each operation. Instead a copy of the shared
pointer can be stored locally, together with the increasing version number of
the corresponding hash table (using the method from \autoref{ss:handle}). At the
beginning of each operation, we can use the local version number to make sure
that the local counting pointer still points to the newest table version. If
this is not the case, a new pointer will be acquired. This happens only once per
version of the hash table. The old table will automatically be freed once every
thread has updated its local pointer. Note that counting pointers cannot be
exchanged in a lock-free manner increasing the cost of changing the current
table (using a lock). This lock could be avoided by using a hazard pointer. We did not do this 

\myparagraph{Prevent Concurrent Updates to ensure Consistency (synchronized)} We
propose a simple protocol inspired by read-copy-update protocols
\cite{mckenney1998read}. The thread $t$ triggering the growing operation sets
some global growing flag using a CAS instruction.  A thread $t$ performing a
table access sets a local busy flag when starting an operation.  Then it
inspects the growing flag, if the flag is set, the local flag is unset.  Then
the local thread waits for the completion of the growing operation, or helps
with migrating the table depending on the current growing strategy. Thread $t$
waits until all busy flags have been unset at least once before starting the
migration. When the migration is completed, the growing flag is reset, signaling
to the waiting threads that they can safely continue their table-operations.
Because this protocol ensures that no thread is accessing the previous table
after the beginning of the migration, it can be freed without using reference
counting.

We call this method (semi-)synchronized, because grow and update operations are
disjoint. Threads participating in one growing step still arrive asynchronously,
e.g.~when the parent application called a hash table operation.  Compared to the
marking based protocol, we save cost during migration by avoiding CAS
operations.  However, this is at the expense of setting the busy flags for
\emph{every} operation. Our experiments indicates that overall this is only
advantageous for updates using atomic operations like fetch-and-add that cannot
coexist with the marker flags.

\subsection{Deletions}\label{ss:delete}

For concurrent linear probing, we combine \emph{tombstoning} (see
\autoref{sec:nongrow}) with our migration algorithm to clean the table once it
is filled with too many \emph{tombstones}. 

A \emph{tombstone} is an element, that has a \texttt{del\_key} in place of its
key. The key $x$ of a deleted entry $\langle x,a\rangle$ is atomically changed
to $\langle\texttt{del\_key}, a\rangle$. Other table operations scan over these
deleted elements like over any other nonempty entry. No inconsistencies can
arise from deletions. In particular, a concurrent find-operations with a torn
read will return the element before the deletion since the delete-operation will
leave the value-slot $a$ untouched. A concurrent insert $\langle x,b\rangle$ might read the
key $x$ before it is overwritten by the deletion and return \Id{false} because
it concludes that an element with key $x$ is already present. This is consistent
with the outcome when the insertion is performed before the deletion in a
linearization.

This method of deletion can easily be implemented in the folklore solution from 
\autoref{sec:nongrow}. But the starting capacity has to be set dependent on the
number of overall insertions, since this form of deletion does not free up any
of the deleted cells. Even worse, tombstones will fill up the table and slow
down find queries.

Both of these problems can be solved by migrating all non-tombstone elements into
a new table.  The decision when to migrate the table should be made solely based
on the number of insertions $I$ ($= \textit{number of nonempty cells}$). The count of all non-deleted elements $I-D$ is then used
to decide whether the table should grow, keep the same size (notice
$\gamma=1$ is a special case for our optimized migration), or shrink. Either way,
all tombstones can be removed in the course of the element migration.

\subsection{Bulk Operations}\label{ss:bulk}

Building a hash table for $n$ elements passed to the constructor can be
parallelized using integer sorting by the hash function value.  This works in
time $O(n/p)$ regardless how many times an element is inserted, i.e., sorting
circumvents contention. 
See the work of Müller et al.\cite{muller2015cache} for a discussion of this phenomenon
in the context of aggregation.

This can be generalized for processing batches of size $m=\Om{n}$ that may even
contain a mix of insertions, deletions, and updates.  We outline a simple
algorithm for bulk-insertion that works without explicit sorting albeit does not
avoid contention.  Let $a$ denote the old size of the hash table and $b$ the
number of insertions. Then $a+b$ is an upper bound for the new table size.  If
necessary, grow the table to that size or larger (see below). Finally, in
parallel, insert the new elements.

More generally, processing batches of size $m=\Om{n}$ in a globally synchronized
way can use the same strategy. We outline it for the case of bulk
insertions. Generalization to deletions, updates, or mixed batches is possible:
Integer sort the elements to be inserted by their hash key in expected time
$\Oh{m/p}$. Among elements with the same hash value, remove all but the last.
Then ``merge'' the batch and the hash table into a new hash table (that may have
to be larger to provide space for the new elements).  We can adapt ideas from
parallel merging \cite{HagRue89}.  We co-partition the sorted insertion array
and the hash table into corresponding pieces of size $\Oh{m/p}$. Most of the
work can now be done on these pieces in an embarrassingly parallel way -- each
piece of the insertion array is scanned sequentially by one thread. Consider an
element $\langle x,a\rangle$ and previous insertion position $i$ in the table.  Then we start
looking for a free cell at position $\max(h(x),i)$


\subsection{Restoring the Full Key Space}\label{ss:restoring_key_space}
Our table uses special keys, like the empty key (\texttt{empty\_key}) and the deleted key
(\texttt{del\_key}). Elements that actually have these keys cannot be stored in
the hash table. This can easily be fixed by using two special slots in the
global hash table data structure.  This makes some case distinction necessary
but should have rather low impact on the overall performance.

One of our growing variants (asynchronous) uses a marker bit in its key field. This halves the
possible key space from $2^{64}$ to $2^{63}$. To regain the lost key space, we
can store the lost bit implicitly. Instead of using one hash table that holds
all elements, we use the two subtables $t_0$ and $t_1$. The subtable $t_0$ holds
all elements whose key does not have its topmost bit set. While $t_1$ stores all
elements whose key does have the topmost bit set, but instead of storing the
topmost bit explicitly it is removed.

Each element can still be found in constant time, because when looking for a
certain key, it is immediately obvious in which table the corresponding element
will be stored. After choosing the right table, comparing the 63 explicitly
stored bits can uniquely identify the correct element. Notice that both empty
keys have to be stored distinctly (as described above).


\subsection{Complex Key and Value Types}\label{sec:add_string_keys} 

Using CAS instructions to change the content of hash table cells makes our data
structure fast but limits its use to cases where keys and values fit into memory
words. Lifting this restriction is bound to have some impact on performance but
we want to outline ways to keep this penalty small. The general idea is to
replace the keys and or values by references to the actual data.

\paragraph*{Complex Keys} To make things more concrete we outline a way where
the keys are strings and the hash table data structure itself manages space for
the keys. When an element $\langle s,a\rangle$ is inserted, space for string $s$
is allocated. The hash table stores $\langle r,a\rangle$ where $r$ is a pointer
to $s$. Unfortunately, we get a considerable performance penalty during table
operations because looking for an element with a given key now has to follow
this indirection for every key comparison -- effectively destroying the
advantage of linear probing over other hashing schemes with respect to cache
efficiency. This overhead can be reduced by two measures: First, we can make the
table bigger thus reducing the necessary search distance -- considering that the
keys are large anyway, this has a relatively small impact on overall storage
consumption. A more sophisticated idea is to store a \emph{signature} of the key
in some unused bits of the reference to the key (on modern machines keys
actually only use $48\,$bits). This signature can be obtained from the master
hash function $h$ extracting bits that were \emph{not} used for finding the
position in the table (i.e. the least significant digits). While searching for a
key $y$ one can then first compare the signatures before actually making a full
key comparison that involves a costly pointer dereference.

Deletions do \emph{not} immediately deallocate the space for the key because
concurrent operations might still be scanning through them.  The space for
deleted keys can be reclaimed when the array grows.  At that time, our migration
protocols make sure that no concurrent table operations are going on.

The memory management is challenging since we need high throughput allocation
for very fine grained variable sized objects and a kind of garbage
collection. On the positive side, we can find all the pointers to the strings
using the hash function. All in all, these properties might be sufficiently
unique that a carefully designed special purpose implementation is faster than
currently available general purpose allocators. We outline one such approach:
New strings are allocated into memory pages of size $\Om{p}$.  Each thread has
one current page that is only used locally for allocating short strings. Long
strings are allocated using a general purpose allocator.  When the local page of
a thread is full, the thread allocates a fresh page and remembers the old one on a
stack. During a shrinking phase, a garbage collection is done on the string
memory. This can be parallelized on a page by page basis. Each thread works on
two pages $A$ and $B$ at a time where $A$ is a partially filled page. $B$ is
scanned and the strings stored there are moved to $A$ (updating their pointer in
the hash table). When $A$ runs full, $B$ replaces $A$. When $B$ runs empty, it is
freed. In either case, an unprocessed page is obtained to become $B$.

\paragraph*{Complex Values}  We can take a similar approach as for complex
keys -- the hash table data structure itself allocates space for complex
values. This space is only deallocated during migration/cleanup phases that make
sure that no concurrent table operations are affected.  The find-operation only
hands out \emph{copies} of the values so that there is no danger of stale
data. There are now two types of update operations. One that modifies part of a
complex value using an atomic CAS operation and one that allocates an entirely
new value object and performs the update by atomically setting the
value-reference to the new object. Unfortunately it is not possible to use both
types concurrently.

\paragraph*{Complex Keys \emph{and} Values}  Of course we can combine the
two approaches described above.  However in that case, it will be more efficient to store a
single reference to a combined key-value object together with a signature.

\section{Using Hardware Memory Transactions}\label{sec:add_tsx}

The biggest difference between 
a concurrent table,
and a sequential hash table is the use of atomic processor instructions. We use
them for accessing and modifying data which is shared between threads. An
additional way to achieve atomicity is the use of hardware transactional memory
synchronization introduced recently by Intel and IBM. The new instruction
extensions can group many memory accesses into a single transaction. All changes
from one transaction are committed at the same time. For other threads they
appear to be atomic.  General purpose memory transactions do not have progress
guarantees (i.e. can always be aborted), therefore they require a fall-back path
implementing atomicity (a lock or an implementation using traditional atomic
instructions).

We believe that transactional memory synchronization is an important opportunity
for concurrent data structures. Therefore, we analyze how to efficiently use
memory transactions for our concurrent linear probing hash tables. In the following,
we discuss which aspects of our hash table can be improved by using restricted
transactional memory implemented in Intel Transactional Synchronization
Extensions (Intel TSX).

We use Intel TSX by wrapping sequential code into a memory transaction. Since
the sequential code is simpler (e.g. less branches, more freedom for compiler
optimizations) it can outperform inherently more complex code based on
(expensive 128-bit CAS) atomic instructions.  As a transaction fall-back
mechanism, we employ our atomic variants of hash table operations.  Replacing the
insert and update functions of our specialized growing hash table with Intel TSX
variants increases the throughput of our hash table by up to 28 \% (see 
\autoref{sec:exp_results}). Speedups like this are easy to obtain on workloads
without contentious accesses (simultaneous write accesses on the same
cell). Contentious write accesses lead to transaction aborts which have a higher
latency than the failure of a CAS. Our atomic fall-back minimizes the penalty
for such scenarios compared to the classic lock-based fall-back that causes more
overhead and serialization.

Another aspect that can be improved through the use of memory transactions is
the key and value size. On current x86 hardware, there is no atomic instruction
that can change words bigger than 128 bits at once. The amount of memory that
can be manipulated during one memory transaction can be far greater than 128
bits. Therefore, one could easily implement hash tables with complex keys and
values using transactional memory synchronization. However, using atomic
functions as fall-back will not be possible. Solutions with fine-grained locks
that are only needed when the transactions actually fail, are still possible.

With general purpose memory transactions it is even possible to atomically
change multiple values that are not stored consecutively. Therefore, it is
possible to implement a hash table that separates the keys from the values
storing each in a separate table. In theory this could improve the cache
locality of linear probing.

Overall, transactional memory synchronization can be used to improve performance
and to make the data structure more flexible.

\section{Implementation Details}\label{sec:impl}

\myparagraph{Bounded Hash Tables.}  All of our implementations are constructed
around a highly optimized variant of the circular bounded \emph{folklore} hash
table that was describe in \autoref{sec:nongrow}.  The main performance
optimizations were to restrict the table size to powers of two -- replacing
expensive modulo operations with fast bit operations
When initializing the capacity $c$, we compute the lowest power of
two, that is still at least twice as large as the expected number of insertions
($2n \leq \textit{size} \leq 4n$).

We also built a second non growing hash table variant called \emph{tsxfolklore},
this variant forgoes the usual CAS-operations that are used to change
cells. Instead tsxfolklore uses TSX transactions to change elements in the table
atomically. As described in \autoref{sec:add_tsx}, we use our usual atomic
operations as fallback in case a TSX transaction is aborted.

\myparagraph{Growing Hash Tables.}  All of our growing hash tables use folklore
or tsxfolklore to represent the current status of the hash table. When the table
is approximately 60\% filled, a migration is started. With each migration, we
double the capacity. The migration works in cell-blocks of the size 4096. Blocks are
migrated with a minimum amount of atomics by using the cluster migration described in
\autoref{ss:growShrink}.


We use a user-space memory pool from Intel's TBB library to prevent a slow down
due to the re-mapping of virtual to physical memory (protected by a coarse lock
in the Linux kernel). This improves the performance of our growing variants,
especially when using more than 24 threads. By allocating memory from this
memory pool, we ensure that the virtual memory that we receive is already mapped
to physical memory, bypassing the kernel lock.

In \autoref{ss:growShrink} we identified two orthogonal problems that have to be
solved to migrate hash tables: which threads execute the migration? and how can
we make sure that copied elements cannot be changed? For each of these problems
we formulated two strategies. The table can either be migrated by user-threads
that execute operations on the table (\emph{u}), or by using a pool of threads
which is only responsible for the migration (\emph{p}). To ensure that copied
elements cannot be changed, we proposed to wait for each currently running
operation synchronizing update and growing phases (\emph{s}), or to mark
elements before they are copied, thus proceeding fully asynchronously
(\emph{a}).

All strategies can be combined -- creating the following four growing hash table
variants: \emph{uaGrow} uses enslavement of user threads and asynchronous marking for
consistency; \emph{usGrow} also uses user threads threads, but ensures
consistency by synchronizing updates and growing routines;
\emph{paGrow} uses a pool of dedicated migration threads for the migration and asynchronous marking of
migrated entries for consistency; and \emph{psGrow} combines the use of a
dedicated thread pool for migration with the synchronized exclusion
mechanism.

All of these versions can also be instantiated using the TSX based non-growing
table tsxfolklore as a basis.


\section{Experimental Evaluation}\label{sec:exp}
We performed a large number of experiments to investigate the performance of
different concurrent hash tables in a variety of circumstances (an overview over all tested hash tables can be found in \autoref{tab:functionality}).  We begin by
describing the tested competitors (\autoref{sec:exp_variants}, our variants are
introduced in \autoref{sec:impl}), the test instances
(\autoref{sec:exp_instance}), and the test environment
(\autoref{sec:exp_hardware}).  Then \autoref{sec:exp_results} discusses the
actual measurements. In \autoref{ss:price}, we conclude the section by summarizing our experiments and reflecting how different generalizations affect the performance of hash tables.

\subsection{Competitors}\label{sec:exp_variants}
To compare our implementation to the current state of the art we use a broad
selection of other concurrent hash tables.  
These tables were chosen on the basis of their popularity in applications and
academic publications. We split these hash table implementations into the
following three groups depending on their growing functionality.

\subsubsection{Efficiently Growing Hash Tables}
This group contains all hash tables, that are able to grow efficiently from a
very small initial size. They are used in our growing benchmarks, where we
initialize tables with an initial size of $4096$ thus making growing
necessary. 

\paragraph*{Junction Linear \junca, Junction Grampa \juncb, and Junction Leapfrog \juncc}
The junction library consists of three different variants of a dynamic
concurrent hash table. It was published by Jeff Preshing over github
\cite{junctionGit}, after our first publication on the subject
(\cite{hashTRArxiv}).  There are no scientific publications, but on his blog
\cite{preshBlog} Preshing writes some insightful posts on his implementation.
In theory, junction's hash tables use an approach to growing which is similar to
ours.  A filled bounded hash table is migrated into a newly allocated bigger
table.  Although they are constructed from a similar idea, the execution seems
to differ quite significantly.  The junction hash tables use a
\emph{quiescent-state based reclamation} (QSBR) protocol, for memory
reclamation.  Using this protocol, in order to reclaim freed hash table memory,
the user has to regularly call a designated function.

Contrary to other hash tables, we used the provided standard hash function
(avalanche), because junction assumes its hash function, to be
invertible. Therefore, the hash function which is used for all other tables (see
\autoref{sec:exp_instance}) is not usable.

The different hash tables within junction all perform different variants of open
addressing.  These variants are described in more detail, in one of Preshing's
blogposts (see~\cite{preshBlog}).

\paragraph*{tbbHM \tbba and tbbUM \tbbb} (correspond to the TBB maps
\texttt{tbb::concurrent\_hash\_map} and \texttt{tbb::concurrent\_unordered\_map}
respectively) The Threading Building Blocks \cite{TBBCite} (TBB) library
(Version 4.3 Update 6) developed by Intel is one of the most widely used
libraries for shared memory concurrent programming. The two different
concurrent hash tables it contains behave relatively similar in our tests.
Therefore, we sometimes only plot the results of tbbHM \tbba. But they have some
differences concerning the locking of accessed elements. Therefore, they behave
very differently under contention.

\subsubsection{Hash Tables with Limited Growing Capabilities}
This group contains all hash tables that can only grow by a limited amount
(constant factor of the initial size) or become very slow when growing is
required.  When testing their growing capabilities, we usually initialize these
tables with half their target size. 
This is comparable to a workload where the approximate number of elements is
known but cannot be bound strictly.

\paragraph*{folly \folly} (\texttt{folly::AtomicHashMap}) This hash table was
developed at facebook as a part of their open source library
folly~\cite{FollyCite} (Version 57:0). It uses restrictions on key and data
types similar to our folklore implementation. In contrast to our growing
procedure, the folly table grows by allocating additional hash tables. This
increases the cost of future queries and it bounds the total growing factor to
$\approx 18$ ($\times \textit{initial
  size}$).  

\paragraph*{cuckoo \cuck} (\texttt{cuckoohash\_map}) This hash table using (bucket)
cuckoo hashing as its collision resolution method, is part of the small
libcuckoo library (Version 1.0). It uses a fine grained locking approach
presented by Li et al.~\cite{AlgoImpCuckoo14} to ensure consistency. Cuckoo is
mentionable for their interesting interface, which combines easy container style
access with an update routine similar to our update interface. 

\paragraph*{RCU \rcua/RCU QSBR \rcub} 
This hash
table is part of the Userspace RCU library (Version 0.8.7) \cite{lwnRCU}, that brings the read
copy update principle to userspace applications. Read copy update is a set of
protocols for concurrent programming, that are popular in the Linux kernel
community \cite{mckenney1998read}. The hash table uses split-ordered lists to grow in a lock-free manner. This approach has been proposed by Shalev and Shavit \cite{splitOrdered}.

RCU uses the recommended read-copy-update variant called urcu. RCU QSBR uses a
QSBR based protocol that is comparable to the one used by junction hash
tables. It forces the user to repeatedly call a function with each participating
thread. We tested both variants, but in many plots we show only RCU \rcua
because both variants behaved very similarly in our tests.

\subsubsection{Non-Growing Hash Tables}
One of the most important subjects of this publication is offering a
scalable asynchronous migration for the simple folklore hash table. While this
makes it usable in circumstances where bounded tables cannot be used, we want to
show that even when no growing is necessary we can compete against bounded hash
tables. Therefore, it is reasonable to use our growing hash table even in
applications where the number of elements can be bounded in a reasonable manner,
offering a graceful degradation in edge cases and allowing improved memory usage
if the bound is not reached.

\paragraph*{Folklore \folk}
Our implementation of the folklore solution described in
\autoref{sec:nongrow}. Notice that this hash table is the core of our
growing variants. Therefore, we can immediately determine the overhead that the
ability for growing places on this implementation (Overhead for approximate
counting and shared pointers).

\paragraph*{Phase Concurrent \phase}
This hash table implementation proposed by Shun and Blelloch \cite{shun2014phase} is designed to
support only phase concurrent accesses, i.e.~no reads can occur concurrently
with writes.  We tested this table anyway, because several of our test instances
satisfy this constraint and it showed promising running times.

\paragraph*{Hopscotch Hash \hops}
Hopscotch hashing (ver 2.0) is one of the more popular variants of open addressing. The
version we tested, was published by Herlihy et al.~\cite{herlihy2008hopscotch} connected to their original
publication proposing the technique.
Interestingly, the provided implementation only implements the functionality of a
hash set (unable to retrieve/update stored data). Therefore, we had to adapt
some tests to account for that (\texttt{insert}$\cong$\texttt{put} and
\texttt{find}$\cong$\texttt{contains}).

\paragraph*{LeaHash \lea}
This hash table is designed by Lea \cite{leahash} as part of Java's Concurrency
Package.  We have obtained a C++ implementation which was published together
with the hopscotch table. It was previously used during for experiments by 
Herlihy et al.~\cite{herlihy2008hopscotch} and Shun and Blelloch \cite{shun2014phase}. LeaHash uses hashing with
chaining and the implementation that we use has the same hash set interface as
hopscotch.

\vspace{0.2cm} As previously described, we used hash set implementations for
Hopscotch hashing, as well as LeaHash (they were published like this).  They
should easily be convertible into common hash map implementations, without
loosing too much performance, but probably using quite a bit more memory.


\subsubsection{Sequential Variants}
To report absolute speedup numbers, we implemented sequential variants of
growing and fixed size tables.  They do not use any atomic instructions or
similar slowdowns.  They outperform popular choices like google's dense hash map
significantly (80\% increased insert throughput), making them a reasonable
approximation for the optimal sequential performance.


\begin{table*}[ht]
\caption{\label{tab:functionality}Overview over Table Functionalities.}
\small
\centering
\begin{tabular} {lc|p{2.3cm}p{1.7cm}p{2.5cm}p{1.2cm}p{2.5cm}}
\bf name   & \bf plot & \bf std. interface & \bf growing    & \bf atomic updates & \bf deletion & \bf arbitrary types \\
\hline
xyGrow     &        & & & & & \\
\ \ uaGrow & \uag   & using handles   & \centering\ck & \centering\ck  & \centering\ck &     \\
\ \ usGrow & \usg   & \hspace{5mm} '' & \centering\ck & \centering\ck  & \centering\ck &     \\
\ \ paGrow & \pag   & \hspace{5mm} '' & \centering\ck & \centering\ck  & \centering\ck &     \\
\ \ psGrow & \psg   & \hspace{5mm} '' & \centering\ck & \centering\ck  & \centering\ck & \vspace{1mm}\\
Junction   &        & & & & & \\
\ \ linear & \junca & qsbr function   & \centering\ck & only overwrite & \centering\ck & \\
\ \ grampa & \juncb & \hspace{5mm} '' & \centering\ck & \hspace{7mm} ''& \centering\ck & \\
\ \ leapfrog&\juncc & \hspace{5mm} '' & \centering\ck & \hspace{7mm} ''& \centering\ck & \vspace{1mm}\\
TBB        &        & & & & & \\
\ \ hash map & \tbba& \centering\ck   & \centering\ck & \centering\ck  & \centering\ck & \hspace{1cm}\ck \\
\ \ unordered& \tbbb& \centering\ck   & \centering\ck & \centering\ck  & \centering unsafe & \hspace{1cm}\ck \\
\hline
Folly      & \folly & \centering\ck   & const factor  & \centering\ck  &     &     \\
Cuckoo     & \cuck  & \centering\ck   & slow          & \centering\ck  & \centering\ck & \hspace{1cm}\ck \\
RCU        &        & & & & & \\
\ \ urcu   & \rcua  & register thread & very slow     & \centering\ck  & \centering\ck & \hspace{1cm}\ck \\
\ \ qsbr   & \rcub  & qsbr function   & \hspace{5mm}''& \centering\ck  & \centering\ck & \hspace{1cm}\ck \\
\hline
Folklore   & \folk  & \centering\ck   &     & \centering\ck & & \\
Phase      &\phase  & sync phases     &     & only overwrite & \centering\ck & \\
Hopscotch  & \hops  & \centering\ck   &     & set interface  & \centering\ck & \\
Lea Hash   & \lea   & \centering\ck   &     & set interface  & \centering\ck & \\
\end{tabular}
\end{table*}

\subsubsection{Color/Marker Choice}
For practicality reasons, we chose not to print a legend with all of our
figures. Instead, we use this section to explain the color and marker choices
for our plots (see \autoref{sec:exp_variants} and \autoref{tab:functionality}), hopefully making them more readable.

Some of the tested hash tables are part of the same library. In these cases, we
use the same marker, for all hash tables within that library. The different variants of the hash table are then differentiated using the line color (and filling of the marker).


For our own tables, we mostly use \uag and \usg for uaGrow and usGrow
respectively. 

\subsection{Hardware Overview}
\label{sec:exp_hardware}
Most of our Experiments were run on a two socket machine, with Intel Xeon E5-2670
v3 processors (previously codenamed Haswell-EP). Each processor has 12 cores
running at $2.3\,\text{Ghz}$ base frequency. The two sockets are connected by
two Intel QPI-links. Distributed to the two sockets there are $128\,\text{GB}$
of main memory ($64\,\text{GB}$ each).  The processors support Intel
Hyper-Threading, AVX2, and TSX technologies.

This system runs a Ubuntu distribution with the kernel number
3.13.0-91-generic. We compiled all our tests with gcc 5.2.0 -- using
optimization level \texttt{-O3} and the necessary compiler flags
(e.g.~\texttt{-mcx16}, \texttt{-msse4.2} among others).

Additionally we executed some experiments on a 32-core 4-socket Intel Xeon
E5-4640 (SandyBridge-EP) machine, with $512\,\text{GB}$ main memory (using the
same operating system and compiler), to verify our findings, and show improved
scalability even on 4-socket machines.



\subsection{Test Methodology}\label{sec:exp_instance}
Each test measures the time it takes, to execute $10^8$ hash table operations
(\emph{strong scaling}).  Each data point was computed by taking the average, of
five separate execution times.  Different tests use different hash table
operations and key distributions.  The used keys are pre-computed before the
benchmark is started. Each speedup given in this section is computed as the
\emph{absolute speedup} over our hand-optimized sequential hash table.

The work is distributed between threads dynamically. While there is
work to do, threads reserve blocks of 4096 operations to execute (using an
atomic counter). This ensures a minimal amount of work imbalance, making the
measurements less prone to variance.

Two executions of the same test will always use the same input keys.  Most
experiments are performed with uniformly random generated keys (using the
Mersenne twister random number generator \cite{MatNis98}).  Since real world
inputs may have recurring elements, there can be contention which can
potentially lead to performance issues.  To test hash table performance under
contention, we use Zipf's distribution to create skewed key sequences.  Using
the Zipf distribution, the probability for any given key $k$ is
$P(k)=\frac{1}{k^s\cdot H_{N,s}}$, where $H_{N,s}$ is the $N$-th generalized
harmonic number $\sum^N_{k=1}\frac{1}{k^s}$ (normalization factor) and $N$ is
the universe size ($N=10^8$). The exponent $s$ can be altered to regulate the
contention.  We use the Zipf distribution, because it closely models some real
world inputs like natural language, natural size distributions (e.g.~of firms or
internet pages), and even user behavior (\cite{zipfFirms}, \cite{zipfWebCache},
\cite{zipfInternet}).  Notice that key generation is done prior to the benchmark
execution as to not influence the measurements unnecessarily (this is especially
necessary, for skewed inputs).

As a hash function, we use two CRC32C x86 instructions with different seeds, to
generate the upper and lower 32 bits of each hash value. Their hardware
implementation minimizes the computational overhead.

\subsection{Experiments}\label{sec:exp_results}
The most basic functionality of each hash table is inserting and finding
elements.  The performance of many parallel algorithms depends on the
scalability of parallel insertions and finds. Therefore, we begin our
experiments with a thorough investigation into the scalability of these basic
hash table operations.

\paragraph*{Insert Performance}
\begin{figure*}[ht]
 \centering
  \begin{subfigure}{0.49\textwidth}
  \includegraphics[width=\textwidth]{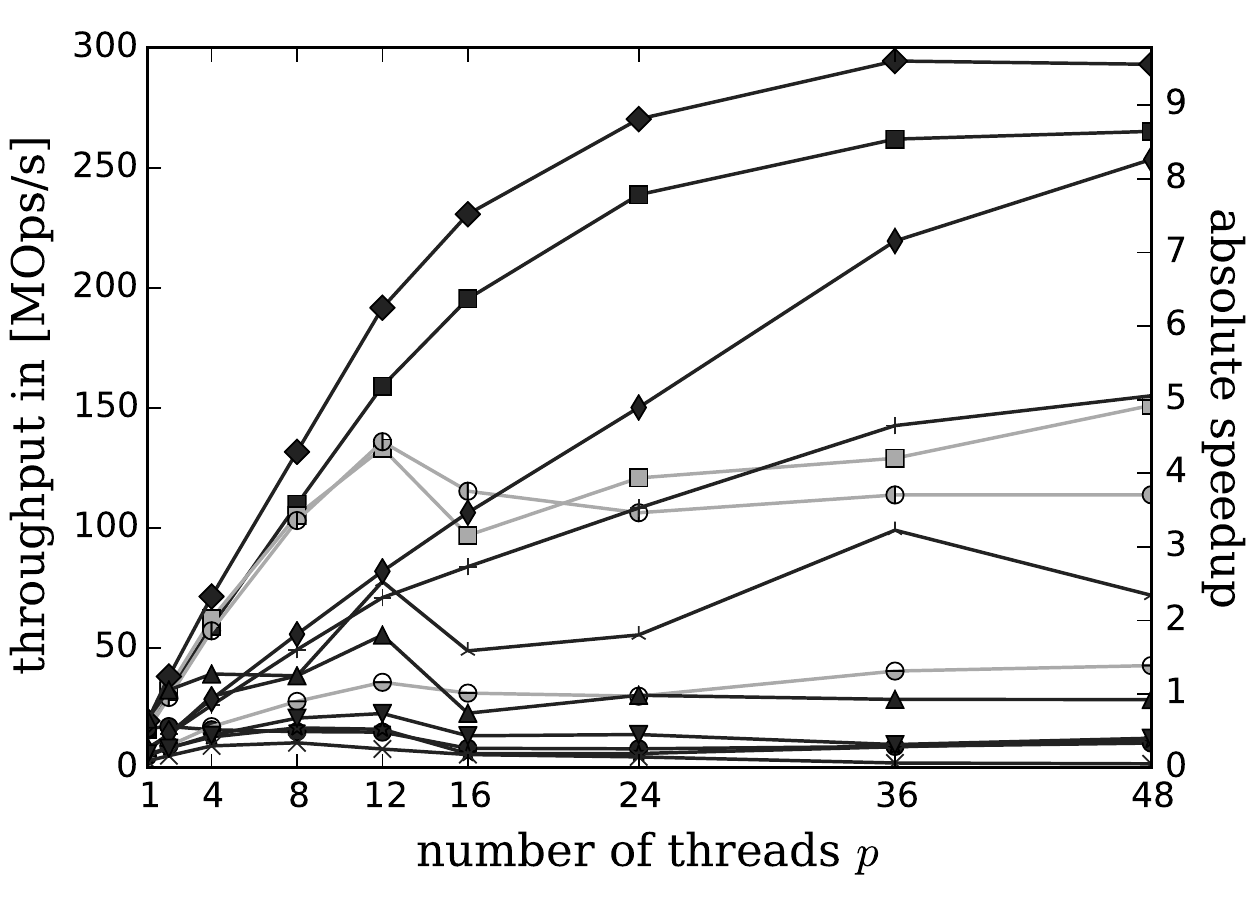}
  \caption{\label{fig:in_prein} Insert into pre-initialized table}
  \end{subfigure}
  \begin{subfigure}{0.49\textwidth}
  \includegraphics[width=\textwidth]{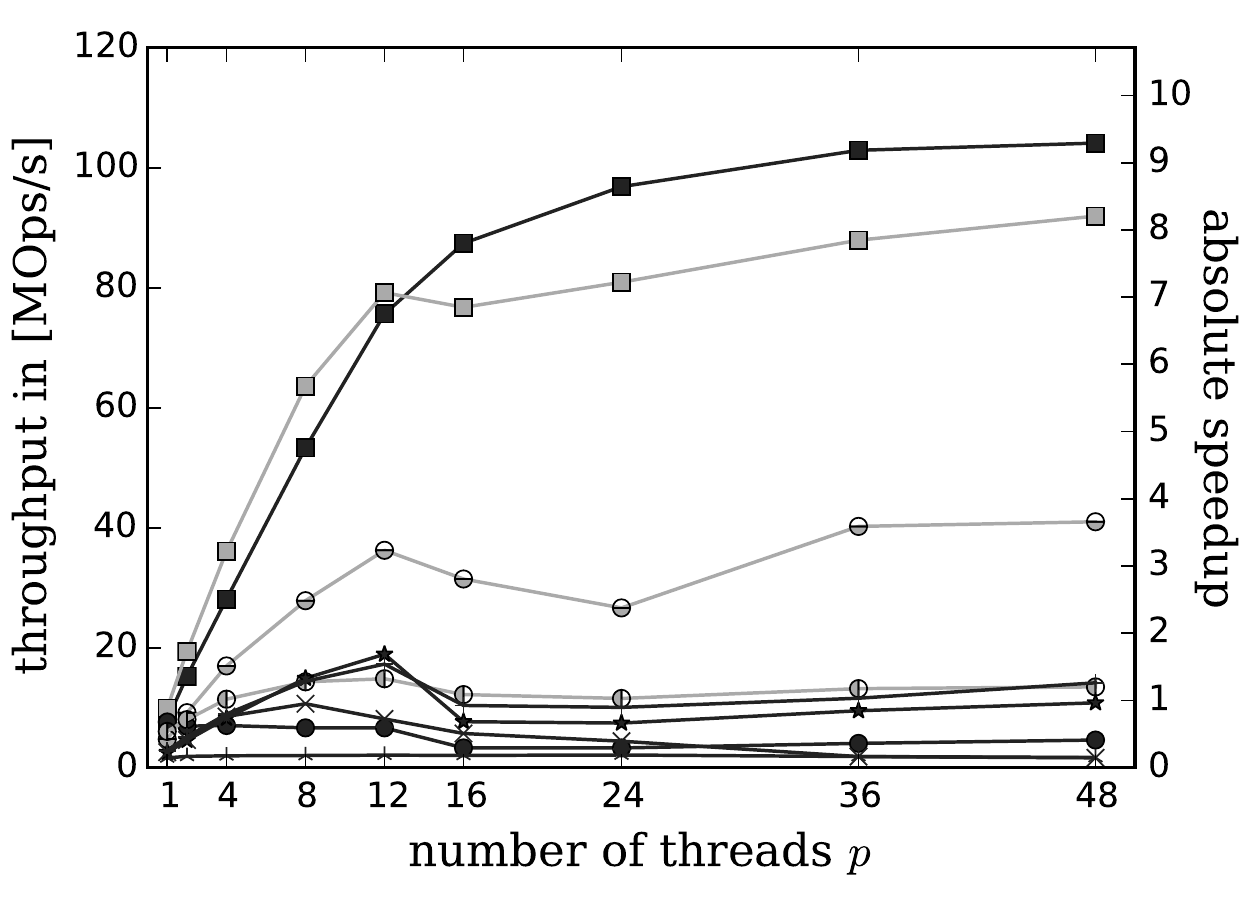}
  \caption{\label{fig:in_grow} Insert into growing table}
  \end{subfigure}
  \caption{\label{fig:in_test} Throughput while inserting $10^8$ elements into a previously empty table (Legend see \autoref{tab:functionality}).}
\end{figure*}
We begin with the very basic test of inserting $10^8$ different uniformly random
keys, into a previously empty hash table. For this first test, all hash tables
have been initialized to the final size making growing unnecessary.  The results
presented in \autoref{fig:in_prein} show clearly, that the folklore \folk
solution is optimal in this case. Since there is no migration necessary, and the
table can be initialized large enough, such that long search distances become
very improbable.  The large discrepancy between the folklore \folk solution, and
all previous growable hash tables is what motivated us, to work with growable
hash tables in the first place.  As shown in the plot, our growing hash table
uaGrow \uag looses about $10\%$ of performance over folklore \folk ($9.6\times$
Speedup vs.~$8.7\times$).  This performance loss can be explained with some
overheads that are necessary for eventually growing the table (e.g.~estimating
the number of elements).  All hash tables that have a reasonable performance
($>50\%$ of folklore~\folk performance), are variants of open addressing
(junction leapfrog \juncc $4.4$ at $p=12$, folly \folly $5.1$, phase \phase
$8.3$) that have similar restrictions on key and value types. All hash tables that can handle generic data types are severely outclassed (\tbba, \tbbb, \cuck, \rcua, and \rcub).

After this introductory experiment, we take a look at the growing
capability of each table.  We again insert $10^8$ elements into a previously empty table. This
time, the table has only been initialized, to hold 4092 elements ($5\cdot10^7$ for
all \emph{semi growing} tables). We can clearly see from the plots in
\autoref{fig:in_grow}, that our hash table variants are significantly faster
than any comparable tables.  The difference becomes especially obvious once two
sockets are used ($>12$ cores). With more than one socket, none of our
competitors could achieve any significant speedups. On the contrary, many tables
become slower when executed on more cores. This effect, does not happen for our
table.

Junction grampa \juncb is the only growing hash table -- apart from our growing
variants -- which achieves absolute speedups higher than $2$. Overall, it is
still severely outperformed by our hash table uaGrow \uag (factor
$2.5\times$). Compared to all other tables, we achieve at least seven times the
performance (descending order; using 48 threads) folly \folly
($7.4\times$), junction leapfrog \juncc ($7.7\times$), tbb hm \tbba ($9.6\times$), tbb um \tbbb
($10.7\times$), junction linear \junca ($22.6\times$), cuckoo \cuck ($61.3\times$), rcu \rcua
($63.2\times$), and rcu with qsbr \rcub ($64.5\times$).


The speedup in this growing instance is even better than the speedup in our
non-growing tests.  Overall we reach absolute speedups of $>9\times$ compared
to the sequential version (also with growing). This is slightly better then the
absolute speedup in the non-growing test ($\approx 8.5$), suggesting that our
migration is at least as scalable as hash table accesses. Overall the insert
performance of our implementation behaves as one would have hoped.  It performs
similar to folklore \folk in the non-growing case, while performing similarly well in
tests where growing is necessary.

\begin{figure*}[ht]
 \centering
  \begin{subfigure}{0.49\textwidth}
  \includegraphics[width=\textwidth]{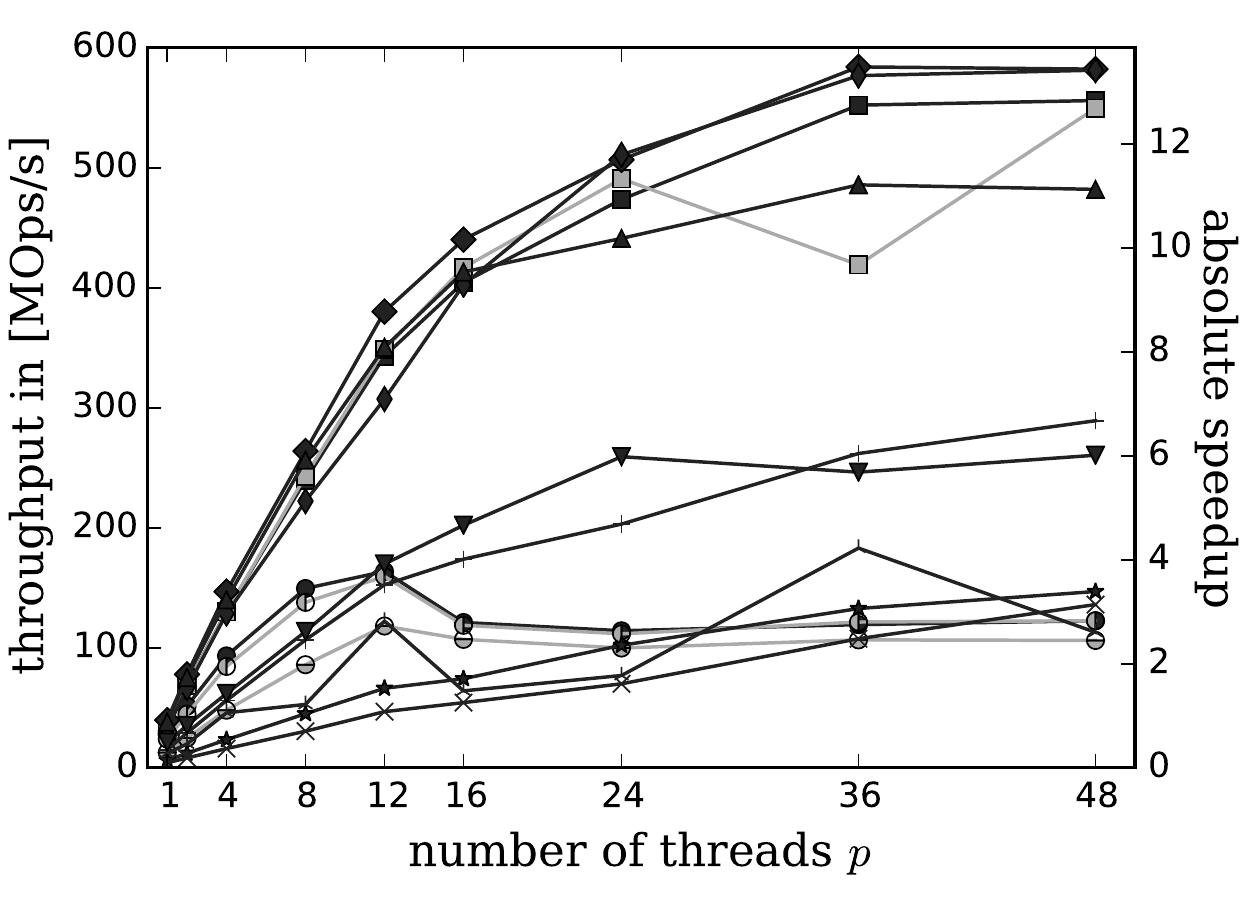}
  \caption{\label{fig:in_find_s} Successful finds}
  \end{subfigure}
  \begin{subfigure}{0.49\textwidth}
  \includegraphics[width=\textwidth]{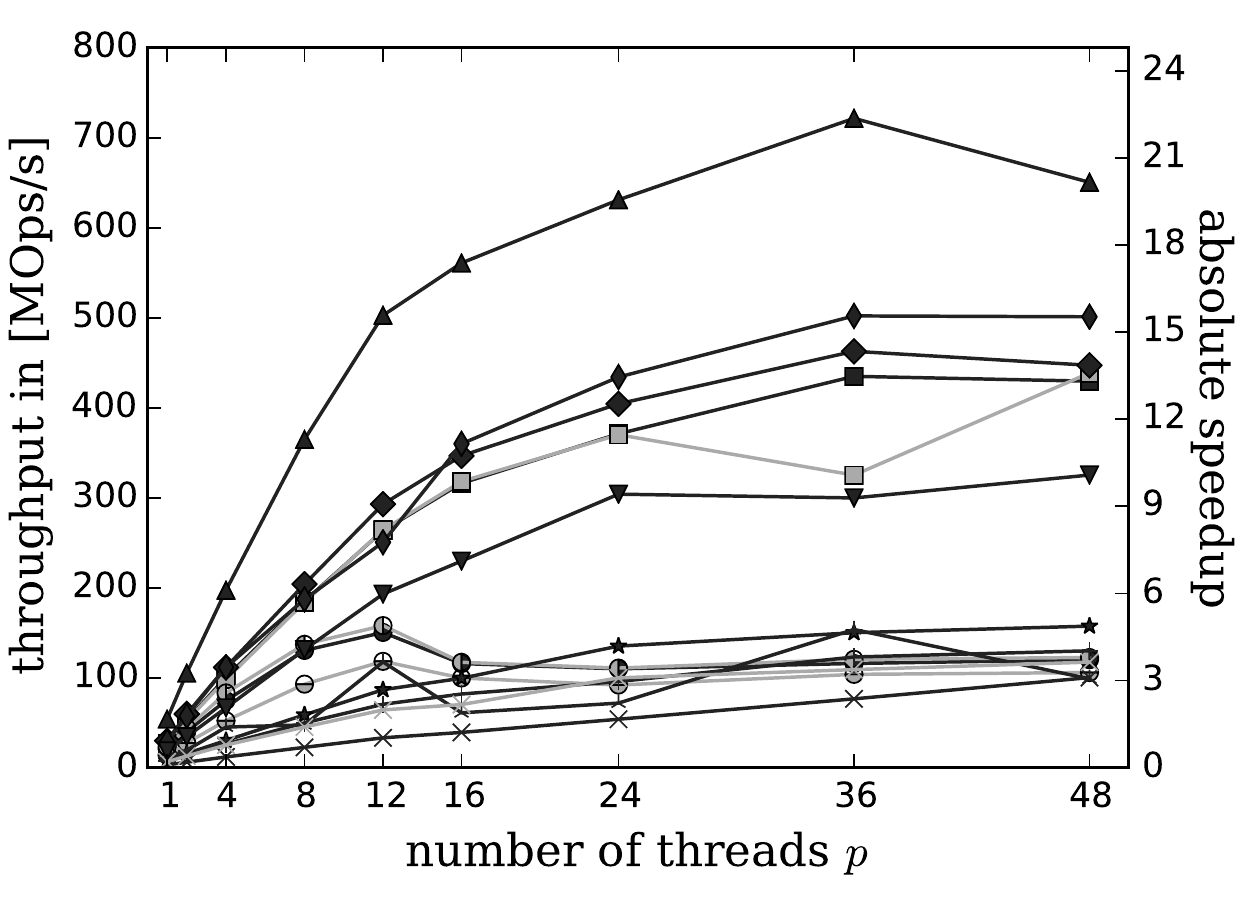}
  \caption{\label{fig:in_find_u} Unsuccessful finds}
  \end{subfigure}
  \caption{\label{fig:in_find} Performance and scalability calling $10^8$ unique find operations, on a table, containing $10^8$ unique keys (Legend see \autoref{tab:functionality}).}
\end{figure*}

\paragraph*{Find Performance}
When looking for a key in a hash table there are two possible outcomes, either it is in the
table or it is not. For most hash tables not finding an element
takes longer than finding said element. Therefore, we present two distinct
measurements for both cases \autoref{fig:in_find_s} and \autoref{fig:in_find_u}.
The measurement for successful finds has been made by looking for $10^8$
elements, that have previously been inserted into a hash table. For the
unsuccessful measurement, $10^8$ uniformly random keys are searched in this same
hash table.

All the measurements made for these plots were done on a preinitialized table
(preinitialized before insertion). This does not make a difference for our
implementation, but it has an influence on some of our competitors. All tables
that grow by allocating additional tables (namely cuckoo \cuck and folly \folly) have
significantly worse find performance on a grown table, as they can have multiple
active tables at the same time (all of them have to be checked).

Obviously, find workloads achieve bigger throughputs than insert heavy workloads
-- no memory is changed and no coordination is necessary between
processors (i.e.~atomic operations). It is interesting that find operations
seem to scale better with multiple processors. Here, our growable implementations
achieve speedups of $12.8$ compared to $9$ in the insertion case.

When comparing the find performance between different tables, we can see that
other implementations with open addressing narrow the gap towards our
implementation. Especially, the hopscotch hashing \hops and the phase concurrent
approach \phase seem to perform well when finding elements. Hopscotch hashing \hops performs
especially well in the unsuccessful case, here it outperforms all other hash
tables, by a significant margin.  However, this has to be taken with a grain of
salt, because the tested implementation only offers the functionality of a hash
set (contains instead of find). Therefore, less memory is needed per element and
more elements can be hashed into one cache line, making lookups significantly
more cache efficient.

\begin{figure*}[ht]
 \centering
  \begin{subfigure}{0.69\textwidth}
  \includegraphics[width=\textwidth]{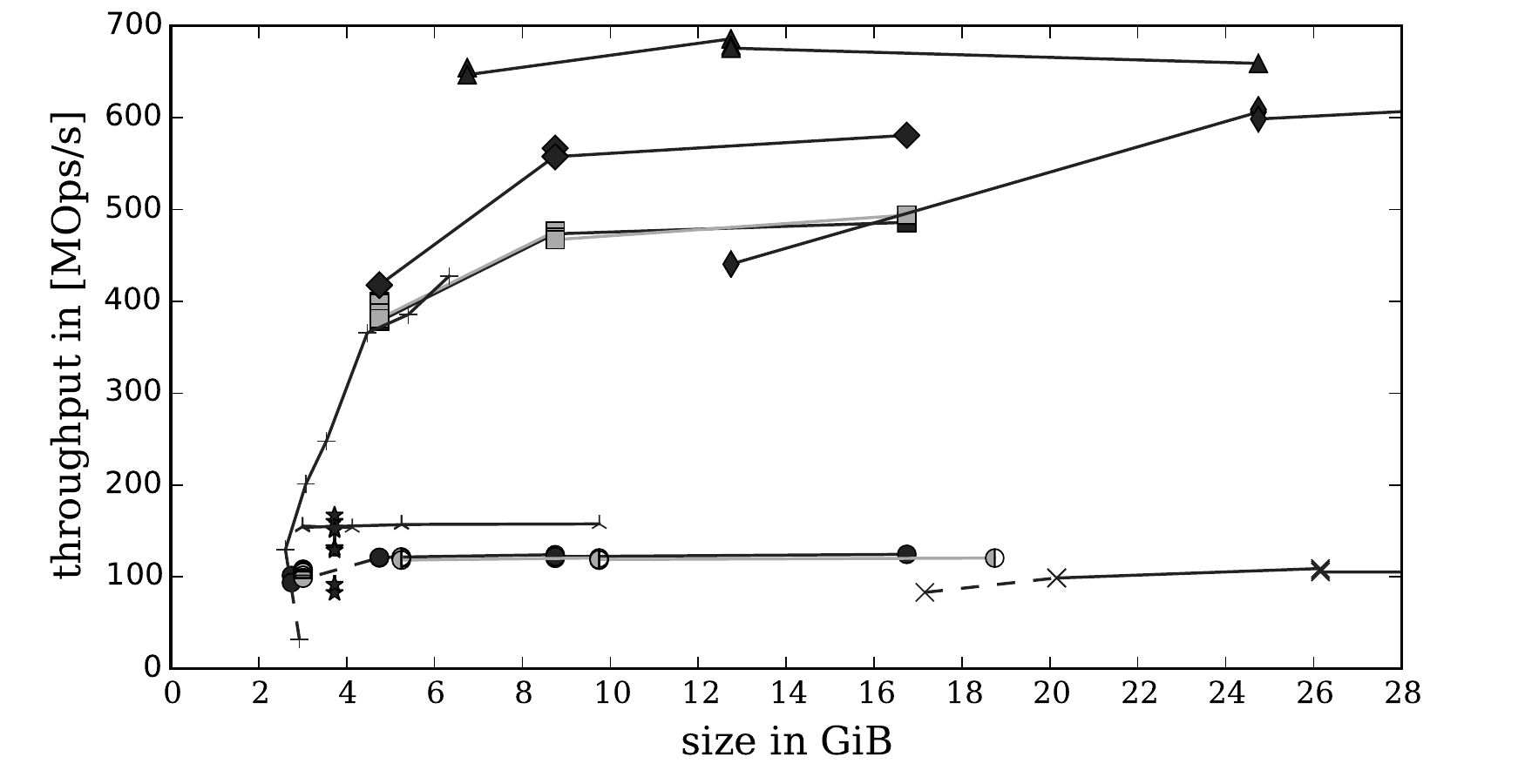}
  \end{subfigure}
  \begin{subfigure}{0.29\textwidth}
    \caption{Performance of unsuccessful find operations
      over the size of the data structure.}
  \end{subfigure}
  \caption{\label{fig:mem_test} For these tests $10^8$ keys are searched
    (unsuccessfully) on a hash table containing $10^8$ elements. Prior to the
    setup of the benchmark, the tables were initialized with different sizes
    (there can be many points on one (x-)coordinate) (Legend see \autoref{tab:functionality}).}
\end{figure*}

For our hash tables, the performance reduction between successful and
unsuccessful finds is around $20$ to $23\,\%$ The difference of absolute speedups
between both cases is relatively small -- suggesting that sequential hash
tables suffer from the same performance penalties.  The biggest difference
has been measured for folly \folly
($51$ to $55\,\%$ reduced performance). Later we see that the reason for this is likely that folly
\folly is configured to use only relatively little memory (see
\autoref{fig:mem_test}).  When initialized with more memory, its performance gets closer to the
performance of other hash tables using open addressing.

\paragraph*{Performance under Contention}
Up to this point, all data sets we looked at contained uniformly random keys
sampled from the whole key space.  This is not necessarily the case in real
world data sets.  For some data sets one keys might appear many times. In some
sets one key might even dominate the input.  Access to this key's element can
slow down the global progress significantly, especially if hash table operations use (fine
grained) locking, to protect hash table accesses.

To benchmark the robustness of the compared hash tables onthese degenerated
inputs, we construct the following test setup.  Before the execution, we compute
a sequence of skewed keys using the Zipf distribution described in
\autoref{sec:exp_instance} ($10^8$ keys from the range $1..10^8$).  Then the
table is filled with all keys from the same range $1..10^8$.

\begin{figure*}[ht]
 \centering
  \begin{subfigure}{0.49\textwidth}
  \includegraphics[width=\textwidth]{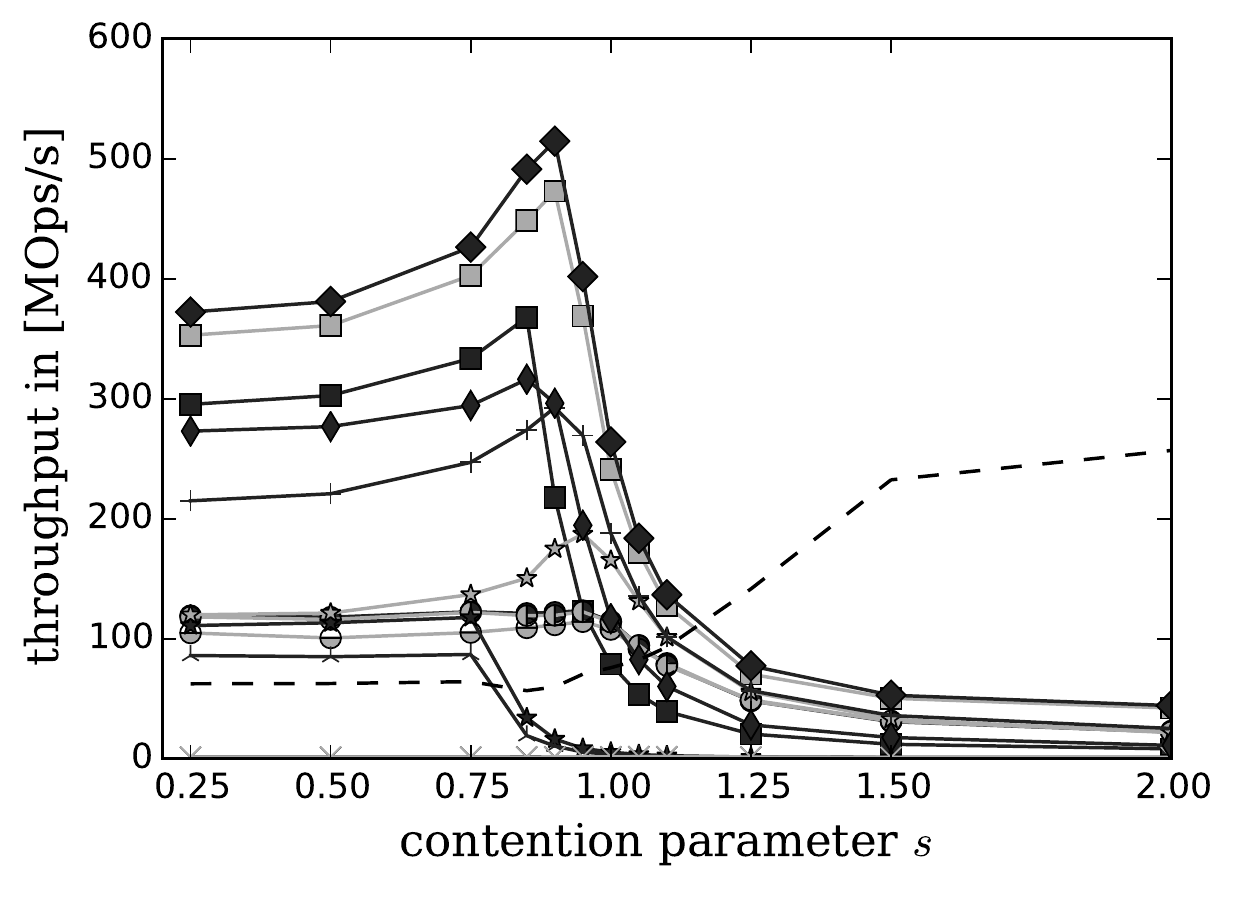}
  \caption{\label{fig:co_upd} Update (Overwrite)}
  \end{subfigure}
  \begin{subfigure}{0.49\textwidth}
  \includegraphics[width=\textwidth]{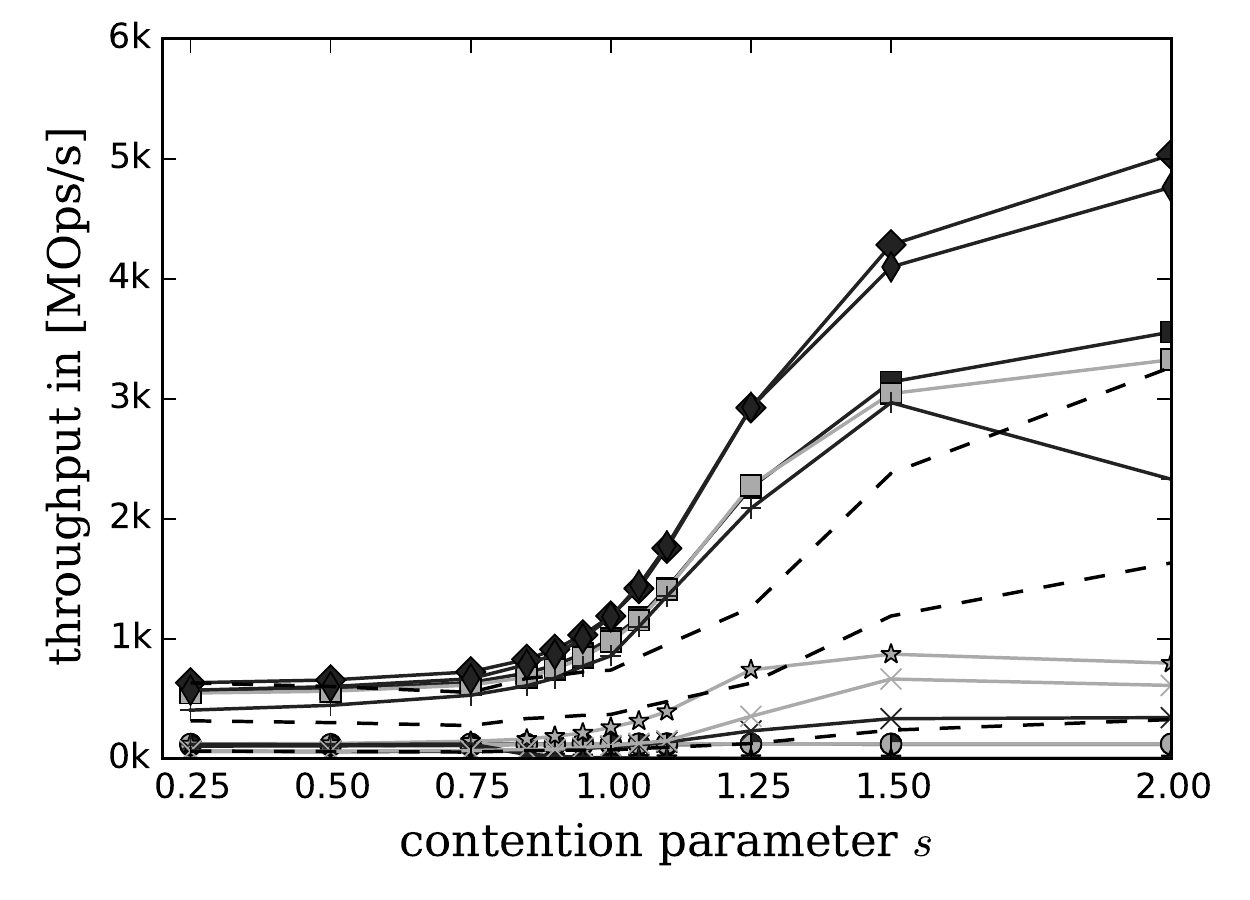}
  \caption{\label{fig:co_find} Successful Finds ($1\times$, $5\times$ and $10\times$ seq.)}
  \end{subfigure}
  \caption{\label{fig:co_test} Throughput executing $10^8$ operations using a
    varying amount of skew in the key sequence (all keys were previously inserted; using 48 threads; Legend see \autoref{tab:functionality}). The sequential performance is indicated using dashed lines.}
\end{figure*}

For the first benchmark we execute an update operation for each key of the
skewed key sequence, overwriting its previously stored element
(\autoref{fig:co_upd}).  These update operations will create contending write
accesses to the hash table. Note that updates perform simple overwrites, i.e.,
the resulting value of the element is not dependent on the previous value.  The
hash table will remain at a constant size for the whole execution, making it
easy to compare different implementations independent of effects introduced
through growing.  In the second benchmark, we execute find operations instead of
updates, thus creating contending read accesses.

For sequential hash tables, contention on some elements can have very positive
effects.  When one cell is visited repeatedly, its contents will be cached and
future accesses will be faster.  The sequential performance is shown in our
figures using a dashed black line.  For concurrent hash tables, contention has 
very different effects.

Unsurprisingly, the effects experienced from contention are different between
writing and reading operations.  The reason is that multiple threads can read
the same value simultaneously, but only one thread at a time can change a value
(on current CPU architecture).  Therefore, read accesses can profit from cache
effects -- much like a sequential hash table, while write accesses are hindered
by the contention.  This goes so far, that for workloads with high contention no
concurrent hash table can achieve the performance of a sequential table.

Appart from slowdown because of exclusive write accesses, there is also the additional problem of cache
invalidation.  When a value is repeatedly changed by different cores of a
multi-socket architecture, then cached copies have to be invalidated whenever
this value is changed. This leads to bad cache efficiency and also to high
traffic on QPI Links (connections between sockets).

From the update measurement shown in \autoref{fig:co_upd} it is clearly visible,
that the serious impact through contention begins between $s = 0.85$ and
$0.95$. Up until that point contention has a positive effect even on update
operations. For a skew between $s = 0.85$ and $0.95$, about $1\,\%$ to $3\,\%$ of
all accesses go to the most common element (key $k_1$). This is exactly the
point where $1/p \approx P(k_1)$, therefore, on average there will be one thread
changing the value of $k_1$.

It is noteworthy that the usGrow \usg version of our hash table is more
efficient when updating than the uaGrow \uag version. The reason for this is
that usGrow uses 128\,bit CAS operations to update elements while simultaneously
making sure, that the marked bit of the element has not been set before the
change.  This can be avoided using the usGrow \usg variant by specializing the
update method to use atomic operations on the data part of the element.  This is
possible because updates and grow routines cannot overlap in this variant.

The plot in \autoref{fig:co_find} shows that concurrent hash tables achieve
performance improvements similar to sequential ones when repeatedly accessing
the same elements.  Our hash table can even increase its speedups over uniform
access patterns, the highest speedup of uaGrow \uag is $17.9$ at $s=1.25$. Since
the speedup is this high, we also included scaled plots showing $5\times$ and
$10\times$ the throughput of the sequential variant. Unfortunately, our growable
variants cannot improve as much, with contention as the non-growing folklore
\folk and phase concurrent \phase tables (both $23.2$ at $s=1.25$). This is
probably due to minor overheads compared to the folklore \folk implementation
which get pronounced since the overall function execution time is reduced.

Overall, we see that our folklore \folk implementation which our growable variants are
based upon, outperforms all other competitors. Our growable variant usGrow \usg is
consistently close to folklore's performance -- outperforming all hash tables that
have the ability to grow.

\paragraph*{Aggregation -- a common Use Case}

Hash tables are often used for key aggregation.  The idea is that all data
elements connected to the same key are aggregated using a commutative and
associative function.  For our test,
we implemented a simple key count program.  To implement the key count routine
with a concurrent hash table, an insert-or-increment function is necessary.  For
some tables, we were not able to implement an update function, where the
resulting value depends on the previous value, within the given interface
(junction tables, rcu tables, phase concurrent, hopscotch, and leahash).  This
was mainly a problem of the used interfaces, therefore, it could probably be
solved by reimplementing a more functional interface.  For our table this can
easily be achieved with the \texttt{insertOrUpdate} interface using an increment as update function (see
\autoref{sec:nongrow}).

The aggregation benchmark uses the same Zipf key distribution as the other contention
tests.  For $10^8$ skewed keys, the insert-or-increment function is
called. Contrary to the previous contention test, there is no pre-initialization.
Therefore, the number of distinct elements in the hash table is dependent on the
contention of the key sequence (given by $s$). This makes growable hash tables
even more desirable, because the final size can only be guessed before the
execution.

\begin{figure*}[ht]
 \centering
  \begin{subfigure}{0.49\textwidth}
  \includegraphics[width=\textwidth]{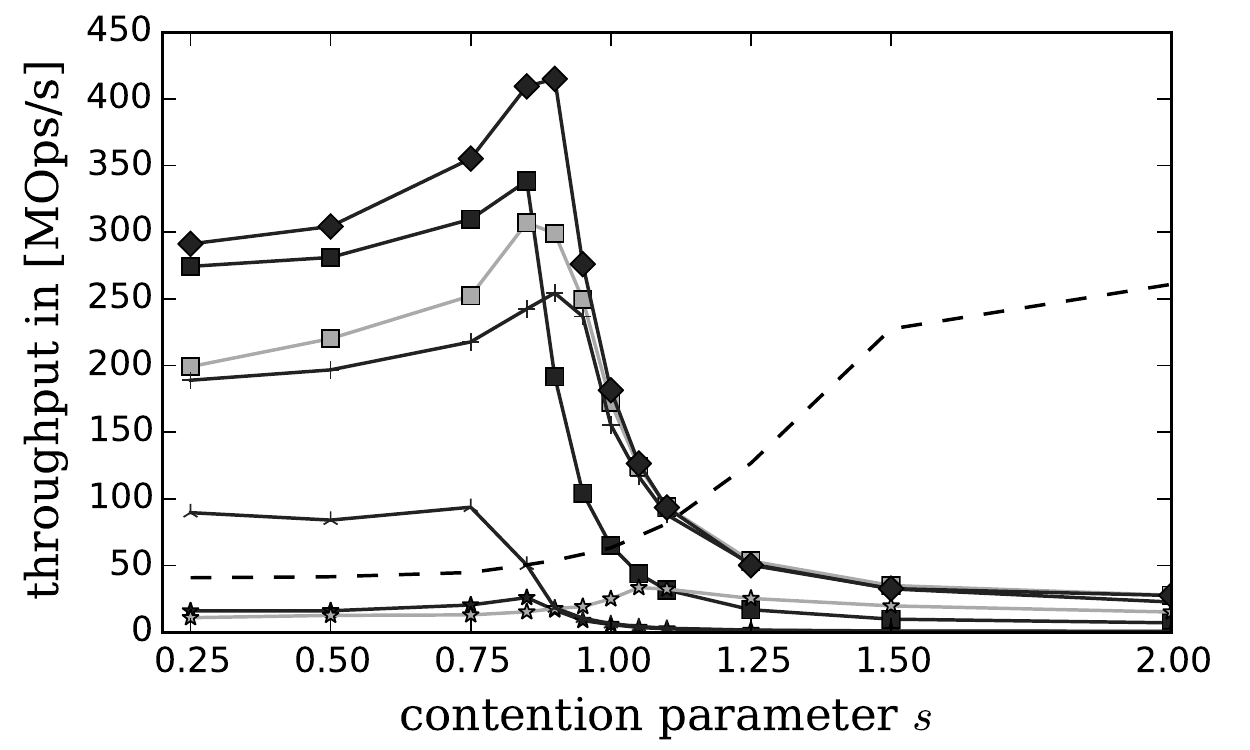}
  \caption{\label{fig:agg_test_f} Aggregation using a pre-initialized size of $10^8$ ($\Rightarrow \textit{size} = |\textit{operations}|$).}
  \end{subfigure}\hspace{1mm}
  \begin{subfigure}{0.49\textwidth}
  \includegraphics[width=\textwidth]{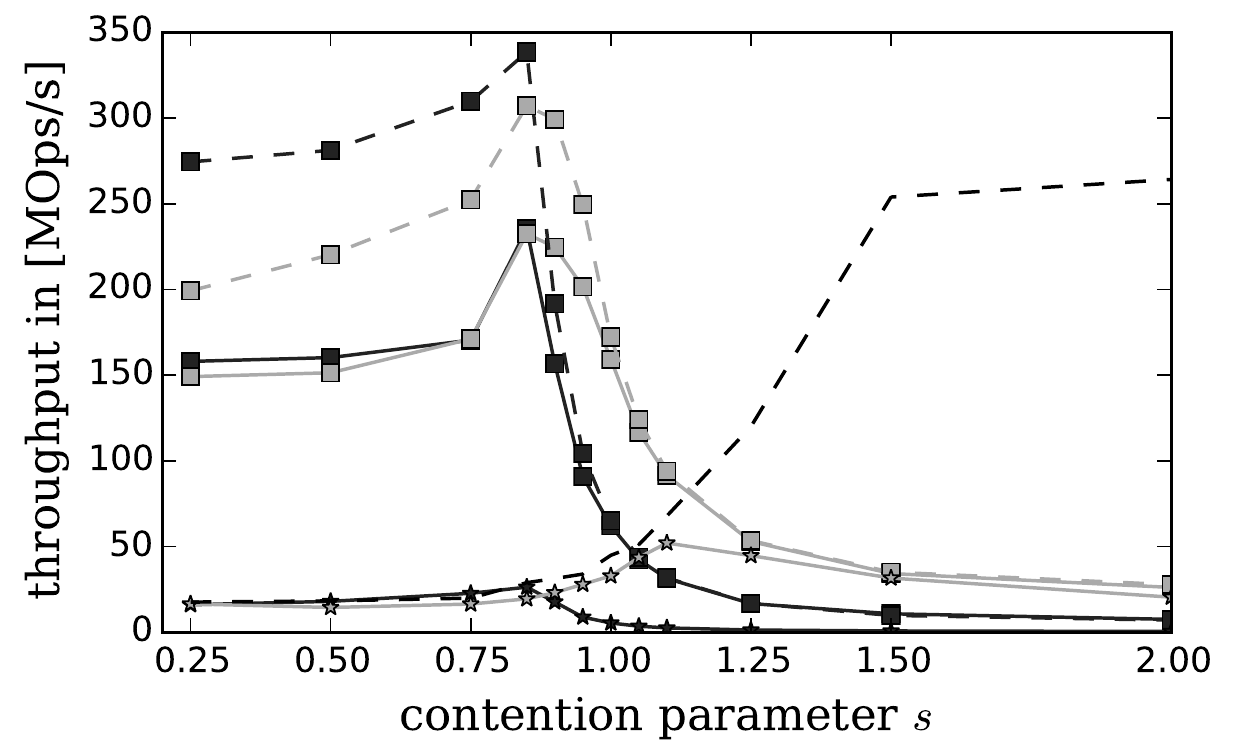}
  \caption{\label{fig:agg_test_u} Aggregation with growing. Dashed plots (\uag and \usg) indicate non-growing performance.}
  \end{subfigure}
  \caption{\label{fig:special_tests} Throughput of an aggregation executing
    $10^8$ insert-or-increment operations using skewed key distributions (using
    48 threads; Legend see \autoref{tab:functionality}).  The dashed black line indicates sequential performance. Some tables
    are not shown because their interface does not support insert-or-increment
    in a convenient way.}
\end{figure*}

Like in previous tests, we make two distinct measurements. One with growing
(\autoref{fig:agg_test_f}) and one without (\autoref{fig:agg_test_u}). In the test
without growing, we initialize the table with a size of $10^8$ to ensure that
there is enough room for all keys, even if they are distinct.  We excluded the
semi-growing tables from \autoref{fig:agg_test_u} as approximating the number of
unique keys can be difficult. To set the growing performance into relation, we
show some non-growing tests. Growing actually costs less in the presence of
contentious updates, because the resulting table will be smaller than without
contention, therefore, fewer growing steps can be amortized over the same number
of operations.

The result of this measurement is clearly related to the result of the
contentious overwrite test shown in \autoref{fig:co_upd}. However, changing a
value by increment has some slight differences to overwriting it, since the
updated value of an insert-or-increment is dependent on its previous value.  In
the best case, this increment can be implemented using an atomic fetch-and-add
operation (i.e. usGrow \usg, folklore \folk, and folly \folly). However this is
not possible for in all hash tables, sometimes dependent updates are implemented
using a read-modify-CAS cycle (i.e. uaGrow \uag) or fine grained locking
(i.e. tbb hash map \tbba or cuckoo \cuck).

Until $s=0.85$, uaGrow \uag seems to be the more efficient option, since it has an
increased writing performance and the update cycle will be successful most of
the time. From that point on, usGrow \usg is clearly more efficient because fetch-and-add behaves better under contention. For highly skewed workloads, it
comes really close to the performance of our folklore implementation \folk which
again performs the best out of all implementations.



\paragraph*{Deletion Tests}
\begin{figure}[ht]
 \centering
  \includegraphics[width=0.5\textwidth]{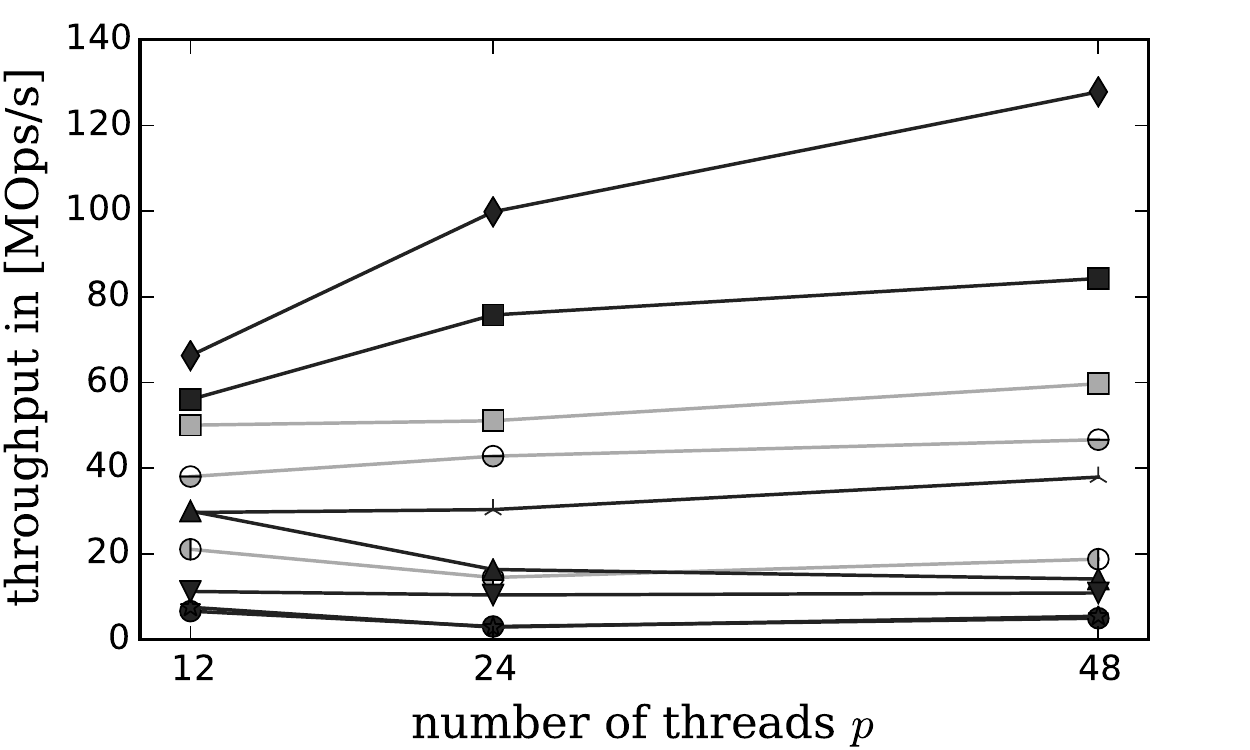}
  \caption{\label{fig:del_test} Throughput in a test using deletion. Some tables
    have been left out of this test, because they do not support deletion with
    memory reclamation (Legend see \autoref{tab:functionality}). By alternating between insertions and
      deletions, we keep the number of elements in the table at approximately
      $10^7$ elements.  For the purpose of computing the throughput, $10^8$ such
      alternations are executed, each counting as one operation
      ($1\,\text{Op} = \text{insert} + \text{delete}$).}
\end{figure}

As described in \autoref{ss:delete}, we use migration, not only to implement an
efficiently growing hash table, but also to clean up the table after
deletions. This way all tombstones are removed, and thus freed cells are
reclaimed. But how does this fare against different ways of removing elements.
This is what we investigate with the following benchmark.

The test starts on a prefilled table ($10^7$ elements) and consists of $10^8$
insertions -- each immediately followed by a deletion. Therefore, the table
remains at approximately the same size throughout the test ($\pm p$ elements).  All
keys used in the test are generated before the benchmark execution (uniform distribution). As described
in \autoref{sec:exp_instance}, all keys are stored in one array.  Each insert
uses an entry from this array distributed in blocks of $4096$ from the
beginning. The corresponding deletion uses the key that is $10^7$ elements prior
to the corresponding insert. The keys stored within the hash table are
contained in a sliding window of the key array.

We constructed the test to keep a constant table size, because this allows us to test non-growing tables without significantly overestimating the necessary
capacity. All hash tables are initialized with $1.5\times 10^7$ capacity,
therefore, it is necessary to reclaim deleted cells to successfully execute the
benchmark.

The measurements shown in \autoref{fig:del_test} indicate, that only the phase
concurrent hash table \phase by Shun and Blelloch \cite{shun2014phase} can outperform our
table.  The reason for this is pretty simple. Their table performs linear
probing comparable to our technique, but it does not use any tombstones for
deletion.  Instead, deleted cells are reclaimed immediately (possibly moving
elements).  This is only possible, because the table does not allow concurrent lookup
operations, thus, removing the possibility for the so called ABA problem (a lookup of
an element while it is deleted returns wrong data, if there is also a concurrent
insert into the newly freed cell).

From all remaining hash tables that support fully concurrent access, ours is
clearly the fastest, even though there are other hash tables like cuckoo \cuck and
hopscotch \hops that also get around full table migrations.


\paragraph*{Mixed Insertions and Finds}

It can be argued that some of our tests are just micro-benchmarks which are not
representative of real world workloads that often mix insertions with lookups.
To address these concerns, we want to show that mixed function workloads
(i.~e.~combined find and insert workloads) behave similarly.

As in previous tests, we generate a key sequence for our test.  Each key of this
sequence is used for an insert or a find operation.  Overall, we
generate $10^8$ keys for our benchmark.  For
each key, insert or find is chosen at random according to the write percentage
$wp$.  In addition to the keys used in the benchmark, we generate a small number
of keys ($pre = 8192\cdot p = 2\,\text{blocks}\cdot p$) that are inserted prior
to the benchmark. This ensures that the table is not empty and there are keys
that can be found with lookups.

The keys used for insertions are drawn uniformly from the key space.  Our goal
for find keys is to pre-construct the find keys in a way that makes find
operations successful and is also fair to all data structures.  If all find
operations were executed to the pre-inserted keys then linear probing hash
tables would have an unfair advantage, because elements that are inserted early
have very short probing distances, while later elements can take much longer to
find.  Therefore any find will look for a random key, that is inserted at least
$8192 \cdot p$ elements earlier in the key sequence.  This key is usually
already in the table when the find operation is called. Looking for a random
inserted element is representative of the overall distribution of probing
distances in the table.

Notice that this method does not strictly ensure that all search keys are
already inserted. In our practical tests we found, that the number of
keys which were not found was negligible for performance purposes (usually below $1000$).

\begin{figure*}[ht]
 \centering
\begin{subfigure}{0.49\textwidth}
  \includegraphics[width=\textwidth]{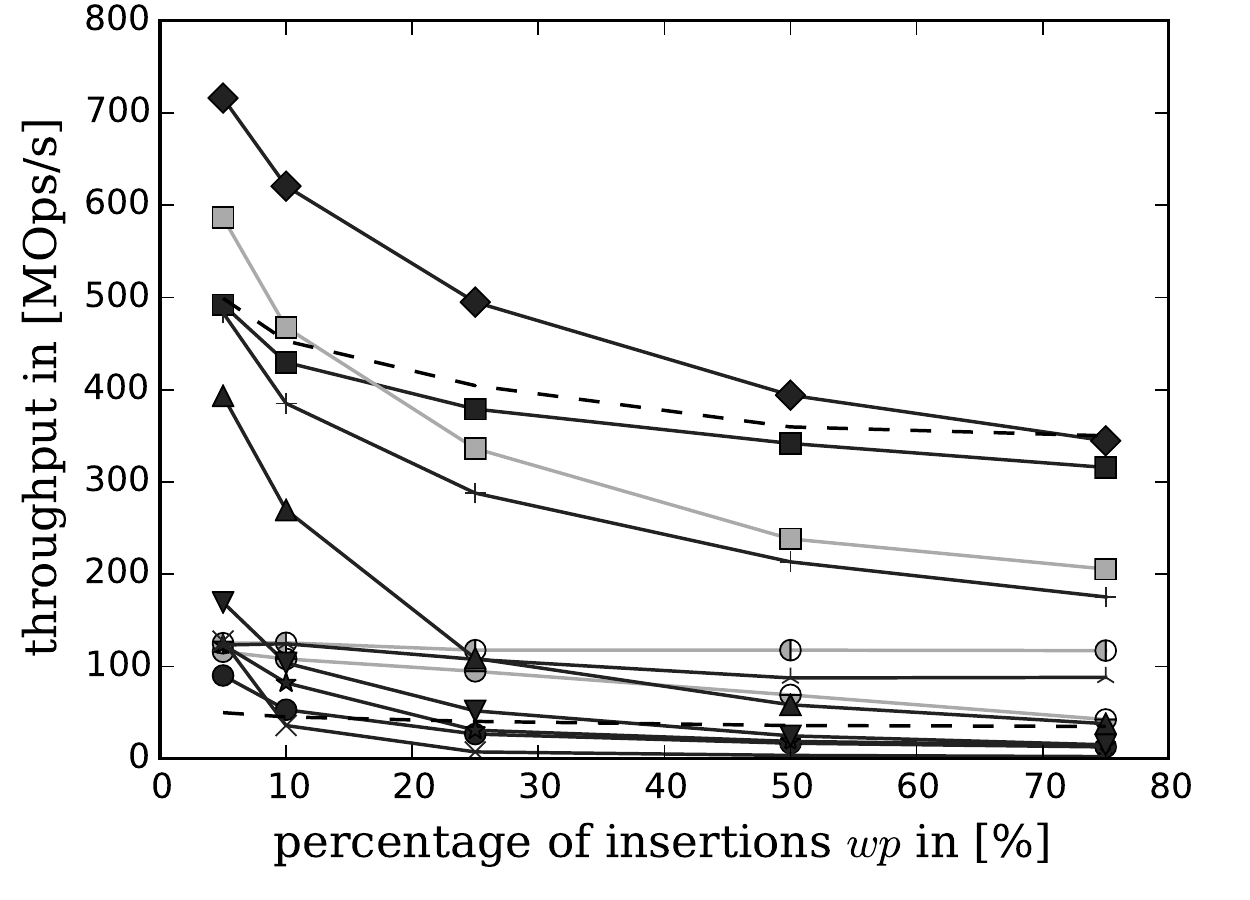}
  \caption{\label{fig:mix_test_ng} Mixed insertions and finds on a pre-initialized table ($\textit{wp}\cdot 10^8+\textit{pre}$).}
  \end{subfigure}
  \begin{subfigure}{0.49\textwidth}
  \includegraphics[width=\textwidth]{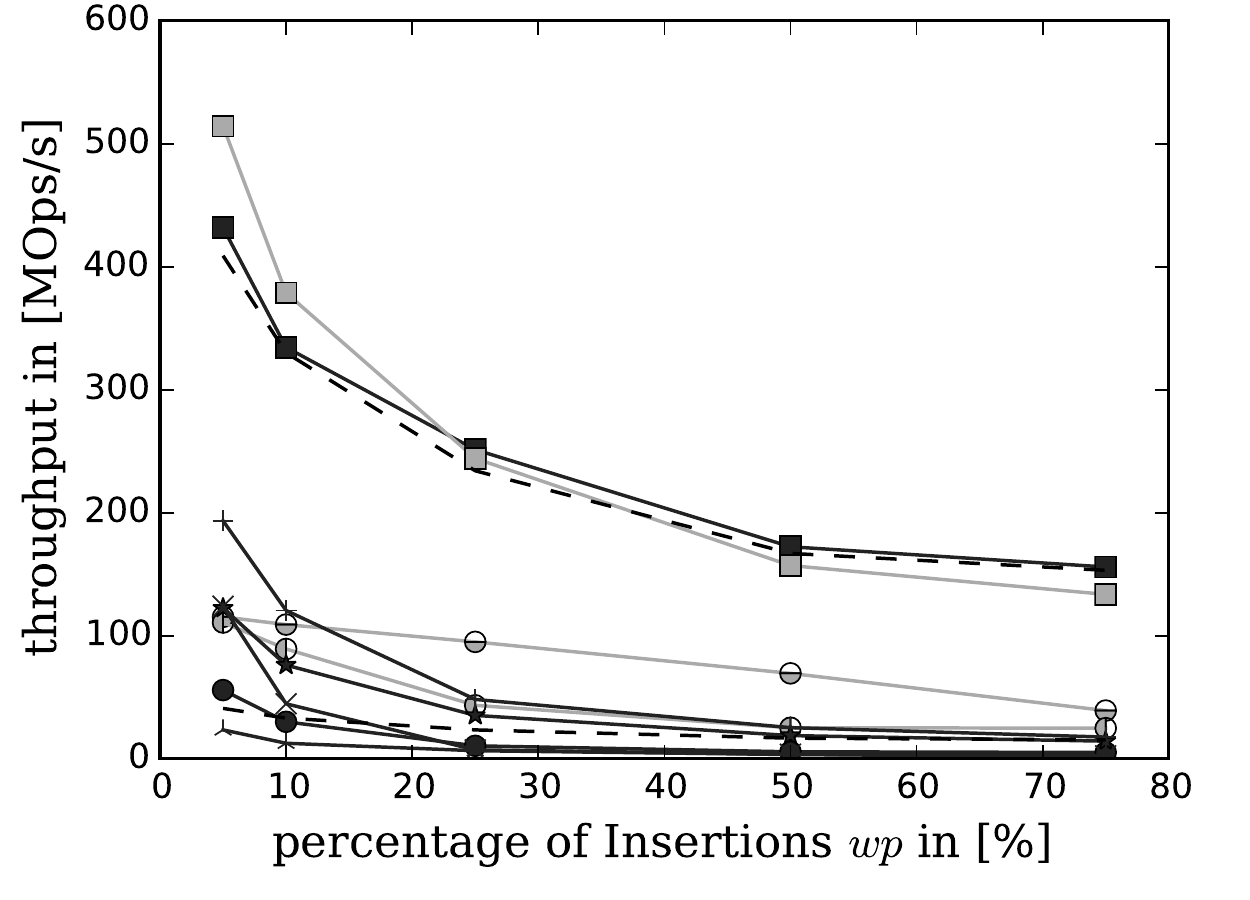}
  \caption{\label{fig:mix_test_g} Mixed insertions and finds on a growing
    table.}
  \end{subfigure}
  \caption{\label{fig:mix_tests} Executing $10^8$ operations mixed between
    insertions and finds (using 48 threads; Legend see \autoref{tab:functionality}).  Dashed lines indicate sequential
    performance ($1\times$ and $10\times$).  Find keys are generated in a fair
    way, that ensures that most find operations are successful.}
\end{figure*}

Comparable to previous tests, we test all hash tables with and without the
necessity to grow the table. In the non-growing test the size of each table is
pre-initialized to be $c = pre + (wp\cdot 10^8)$. In the growing tests
semi-growing hash tables are initialized with half that capacity.

Similar to previous tests it, is obvious that our non-growing linear probing hash
table folklore \folk outperforms most other tables especially on find-heavy
workloads.  Overall, our hash tables behave similar to the sequential solution
with a constant speedup around a factor of $10\times$. Interestingly, the
running time does not seem to be a linear function (over $wp$). Instead,
performance decreases super-linearly. One reason for this could be that for find-heavy workloads, the table remains relatively small for most of the
execution. Therefore, cache effects and similar influences could play a role,
since lookups only look for a small sample of elements that is already in the
table.

\paragraph*{Using Dedicated Growing Threads}

\begin{figure*}[t]
 \centering
  \begin{subfigure}{0.49\textwidth}
  \includegraphics[width=\textwidth]{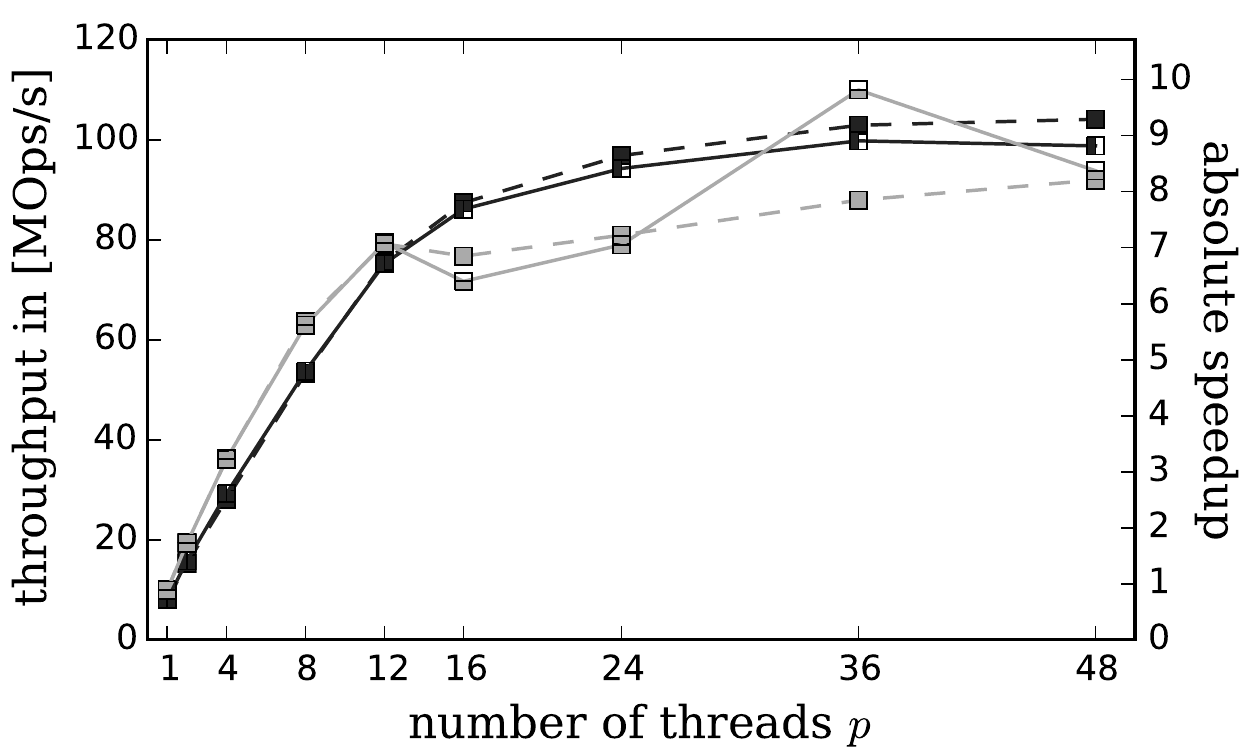}
  \caption{\label{fig:p_ins} Insertions into Growing Table.}
  \end{subfigure}
  \begin{subfigure}{0.49\textwidth}
  \includegraphics[width=\textwidth]{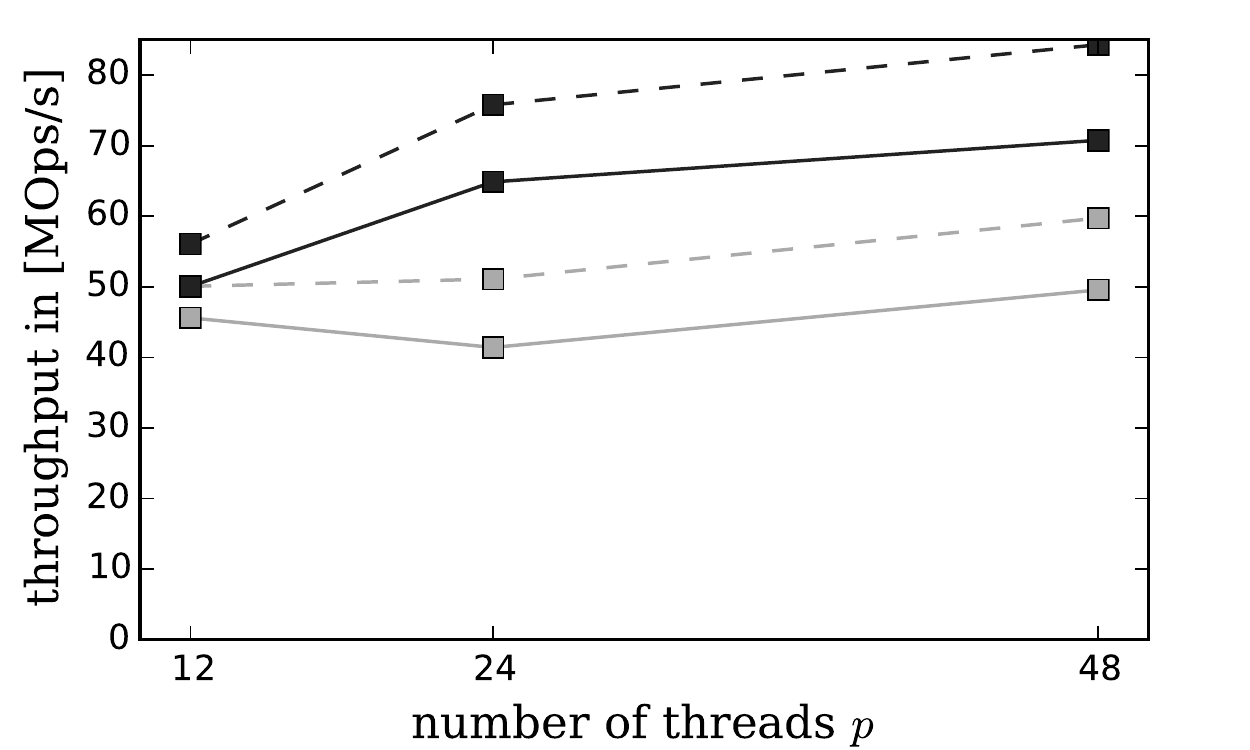}
  \caption{\label{fig:p_del} Alternating Insertions and Deletions.}
  \end{subfigure}
  \caption{\label{fig:thread_grow} Comparison between our regular implementation and the
    variant using a dedicated migration thread pool (dashed lines mark variants enslaving user-threads; Legend see \autoref{tab:functionality}).}
\end{figure*}

In \autoref{ss:async} and \ref{sec:impl} we describe the possibility, to use a
pool of dedicated migration threads which grow the table cooperatively. Usually
the performance of this method does not differ greatly from the performance of
the enslavement variant used throughout our testing.  This can be seen in
\autoref{fig:thread_grow}. Therefore, we omitted these variants from most
plots.

In \autoref{fig:p_ins} one can clearly see the similarities between the the
variants using a thread pool and their counterparts (uaGrow \uag $\cong$ paGrow \pag and usGrow \usg $\cong$ psGrow \psg). The biggest consistent
difference we found between the two options has been measured during the
deletion benchmark in \autoref{fig:p_del}. During this benchmark, insert and
delete are called alternately. This keeps the actual table size constant. For
our implementation, this means that there are frequent migrations on a
relatively small table size. This is difficult when using additional migration
threads, since the threads have to be awoken regularly, introducing some
operating system overhead (scheduling and notification).

\paragraph*{Using Intel TSX Technology}

\begin{figure*}[t]
 \centering
  \begin{subfigure}{0.49\textwidth}
  \includegraphics[width=\textwidth]{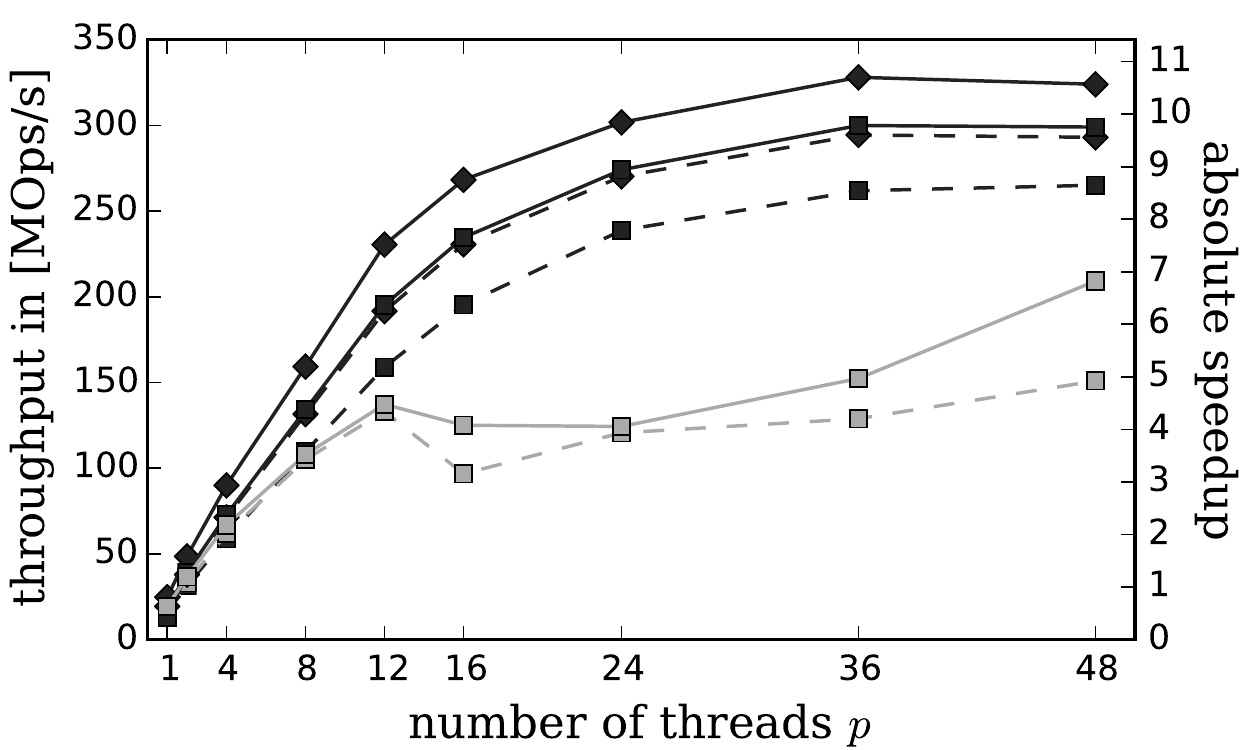}
  \caption{\label{fig:tsx0} Without growing (pre-initialized).}
  \end{subfigure}
  \begin{subfigure}{0.49\textwidth}
  \includegraphics[width=\textwidth]{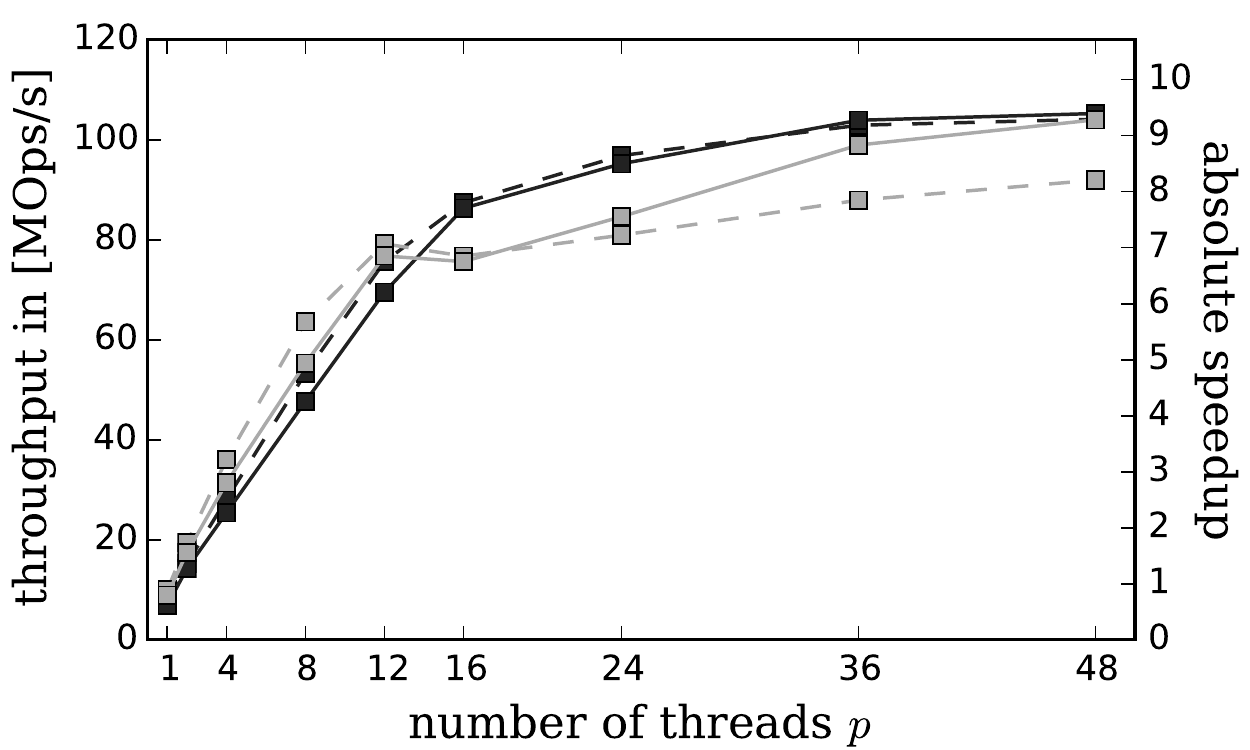}
  \caption{\label{fig:tsx1} Insertions with growing.}
  \end{subfigure}
  \caption{\label{fig:tsx} Comparison between our regular implementation and the
    variant using TSX-transactions in place of atomics (dashed lines are regular variants without TSX).}
\end{figure*}

As described in \autoref{sec:add_tsx}, concurrent linear probing hash tables can
be implemented using Intel TSX technology to reduce the number of atomic
operations. \autoref{fig:tsx} shows some of the results using this approach.

The implementation used in these tests changes only the operations within our
bounded hash table (folklore) to use TSX-transactions.  Atomic fallback
implementations are used, when a transaction fails.  We also instantiated our
growing hash table variants, to use the TSX-optimized table as underlying
hash table implementation.

We tested this variant with a uniform insert workload (see ``Insert
Performance''), because the lookup implementation does not actually need a
transaction. We also show the non-TSX variant, using dashed lines, to indicate the relative performance benefits.

In \autoref{fig:tsx0} one can clearly see that TSX-optimized hash tables offer
improved performance as long, as growing is not necessary. Unfortunately,
\autoref{fig:tsx1} paints a different picture for instances where growing is
necessary. While TSX can be used to improve the usGrow \usg variant of our hash table
especially when using hyperthreading, it offers no performance benefits in the
uaGrow \uag variant. The reason for this is that the running time in these
measurements is dominated by the table migration which is not optimized for
TSX-transactions.

In theory, the migration algorithm can make use of transactions similarly to
single operations. It would be interesting whether an optimized migration could
further improve the growing instances of this test. We have not implemented such
a migration, as it introduces the need for some complex parameter optimizations
-- partitioning the migration into smaller blocks or executing each block-migration
into multiple transactions. We estimate that a well optimized TSX-migration can
gain performance increases on the order of those witnessed in the non-growing
case.

\paragraph*{Memory Consumption}

One aspect of parallel hash tables, that we did not talk about until now
is memory consumption. Overall, a low memory consumption is preferable, but
having less cells means that there will be more hash collisions. This
leads to longer running times especially for non-successful find operations.

Most hash tables do not allow the user to set a specific table size
directly. Instead they are initialized using the expected number of elements.
We use this mechanism to create tables of different sizes.  Using these
different hash tables with different sizes, we find out how well any one hash
table scales when it is given more memory.  This is interesting for applications
where the hash table speed is more important than its memory footprint (lookups
to a small or medium sized hash table within an application's inner loop).

The values presented in the following plot are aquired by initializing the hash
tables with different table capacities
($4096, 0.5\times, 1.0\times, 1.25\times, 1.5\times, 2.0\times, 2.5\times,
3.0\times 10^8$
expected elements; semi- and non-growing hash tables start at $0.5 \times$ and
$1\times$ respectively).  During the test, the memory consumption is measured by
logging the size of each allocation, and deallocation during the execution (done
by replacing allocation methods, e.g.~\texttt{malloc} and
\texttt{memalign}).  Measurements with growing (initial capacity $<10^8$)
are marked with dashed lines.  Afterwards the table is filled with $10^8$
elements. The plotted measurements show the throughput that can be achieved when
doing $10^8$ unsuccessful lookups on the preinitialized table. This throughput
is plotted over the amount of allocated memory each hash table used.

The minimum size for any hash table should be around
$1.53\,\text{GiB} \approx 10^8\cdot (8\,\text{B}+8\,\text{B})$ (Key and Value each have
$8\,\text{B}$).  Our hash table uses a number of cells equal to the smallest
power of $2$ that is at least two times as large as the expected number of
elements.  In this case this means we use $2^{28} \approx 2.7\cdot 10^8$,
therefore, the table will be filled to $\approx 37\,\%$ and use exactly
$4\text{GiB}$. We believe that this memory usage is reasonable, especially for
heavily accessed tables where the performance is important. This is
supported by our measurements as all hash tables that use less memory have bad
performance.

Most hash tables round the number of cells in some convenient
way. Therefore, there are often multiple measurement points using the same
amount of memory.  As expected, using the same amount of memory will usually
achieve a comparable performance. Out of the tested hash tables only the folly \folly
hash table grows linearly with the expected final size. It is also the hash
table, that gains the most performance by increasing its memory. This makes a
lot of sense considering that it uses linear probing and is by default
configured to use more than $50\,\%$ of its cells.

The plot also shows that some hash tables do not gain any performance benefits
from the increased size. Most notable for this are cuckoo \cuck, all variations
of junction \junca \juncb \juncc and the urcu hash tables \rcua. The TBB hash tables \tbba and
\tbbb seem to use a constant amount of memory, independently from the
preinitialized number of elements. This might be a measurement error, caused by
the fact that they use different memory allocation methods (not not logged in our test).

There are also some things that can be learned about growing hash tables from
this plot.  Our migration technique ensures, that our hash table has the exact
same size when growing is required as when it is preinitialized using the same
number of elements. Therefore, lookup operations on the grown table take the
same time as they would on a preinitialized table. This is not true, for many of
our competitors. All Junction tables and RCU produce smaller tables when growing
was used, they also suffer from a minor slowdown, when using lookups on these
smaller tables.  Using Folly is even worse, it produces a bigger table -- when
growing is needed -- and still suffers from significantly worse performance.

\paragraph*{Scalability on a 4-Socket Machine}
Bad performance on multi-socket workloads is recurring theme throughout our
testing. This is especially true for some of our competitors where 2-Socket
running times are often worse than 1-Socket running times. To further expand the
understanding of this problem we made some tests on the 4-Socket machine described in \autoref{sec:exp_hardware}.

\begin{figure*}[ht]
 \centering
  \begin{subfigure}{0.49\textwidth}
  \includegraphics[width=\textwidth]{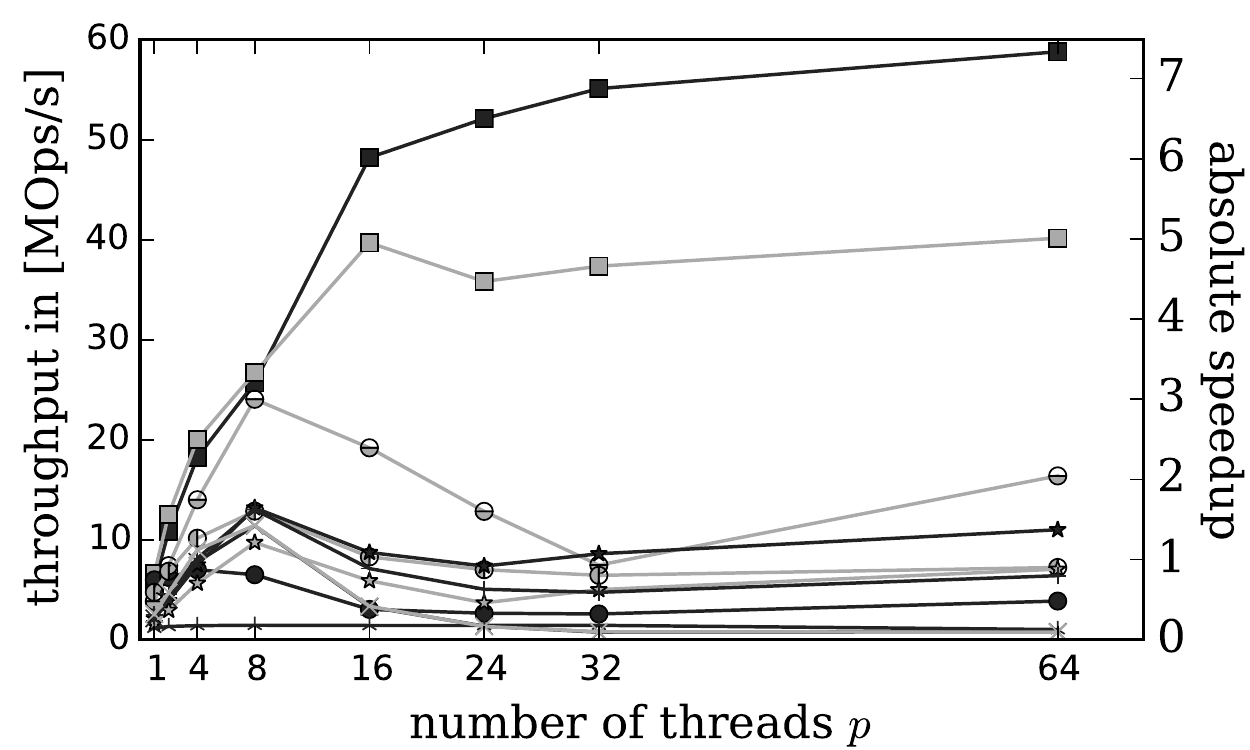}
  \caption{\label{fig:in_127} Insertions into a growing hash table.}
  \end{subfigure}
  \begin{subfigure}{0.49\textwidth}
  \includegraphics[width=\textwidth]{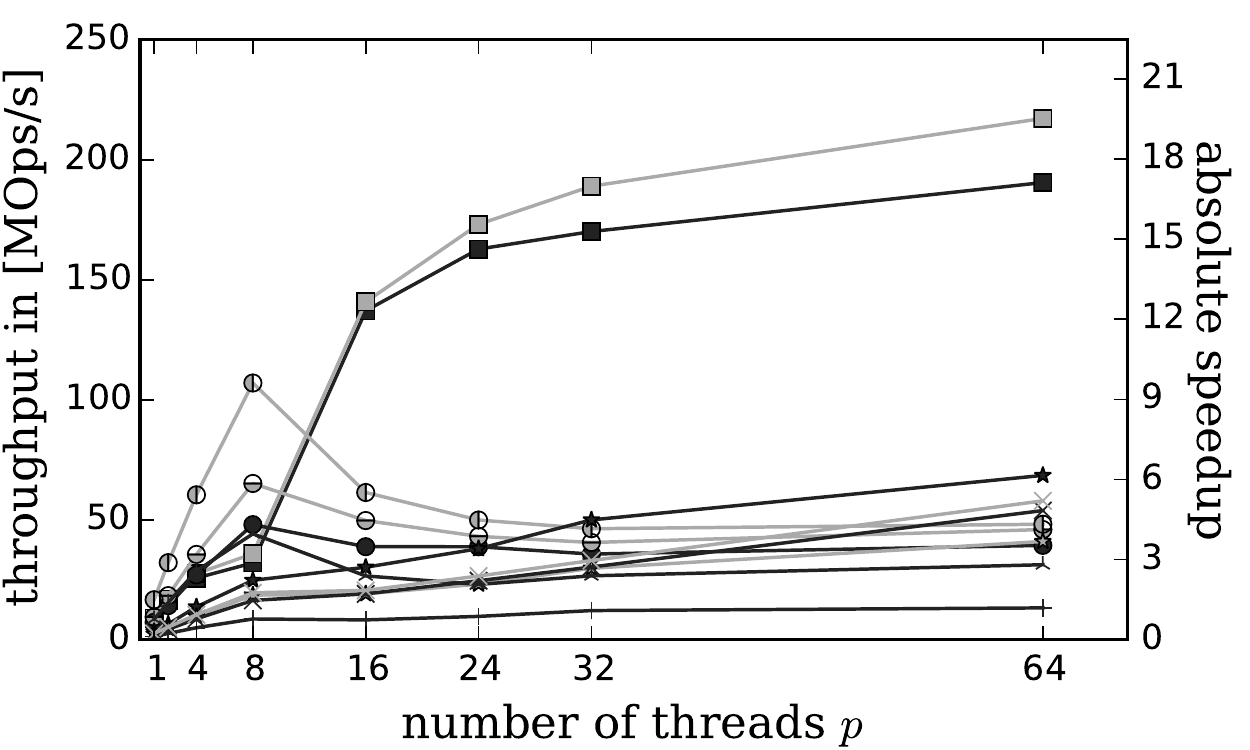}
  \caption{\label{fig:look_127} Unsuccessful Lookups into the filled table.}
  \end{subfigure}
  \caption{\label{fig:127_tests} Basic tests made on our 4-Socket machine,
    consisting of four eight core E5-4640 processors with $2.4\,\text{GHz}$ each
    (codenamed Sandybridge) and $512\,\text{GB}$ main memory  (Legend see \autoref{tab:functionality}).}
\end{figure*}

The used test instances are generated similar to the insert/find tests described
in the beginning of this section ($10^8$ executed operations with uniformly
random keys). The results can be seen in \autoref{fig:in_127} (Insertions) and
\autoref{fig:look_127} (unsuccessful finds).

Our competitor's hash tables
seem to be a lot more effective when using only one of the four sockets
(compared to one of two sockets on the two-socket machine). This is especially
true for the Lookup workload where the junction hash tables \junca start out more
efficient than our implementation. However this effect seems to invert once
multiple sockets are used.

In the test using lookups, there seems to be a performance problem using our hash
table. It seems to scale sub-optimally on one socket. On two sockets
however, the hash table seems to scale significantly better.

Overall the four-socket machine reconfirms our observations. None of our
competitors scale well when a growing hash table is used over multiple
sockets. On the contrary, using multiple sockets will generally reduce the
throughput. This is not the case for our hash table. The efficiency is reduced
when using more then two sockets but the absolute throughput at least remains
stable.

\subsection{The Price of Generality} \label{ss:price}
Having looked at many detailed
measurements, let us now try to get a bigger picture by asking 
which hash table performs well for specific requirements and how much
performance has to be sacrificed for additional flexibility.  This will give us
an intuition, where performance is sacrificed on our way to a fully general hash
table.  Seeing that all tested hash tables fail to scale linearly on
multi-socket machines we try to answer the question if concurrent hash tables
are worth their overhead at all.

At the most restricted level -- no growing/deletions and word sized key and
value types -- we have shown that common linear probing hash tables offer the
best performance (over a number of operations).  Our implementation of this
``folklore'' solution outperforms different approaches consistently, and
performs at least as good as other similar implementations (i.e. the phase
concurrent approach).  We also showed, that this performance can be improved by
using Intel TSX technology. Furthermore, we have shown that our approach to
growing hash tables does not affect the performance on known input
sizes significantly (preinitialized table to the correct size).


Sticking to fixed data types but allowing dynamic growing, the best data
structures are our growing variants
($\{\text{ua}, \text{us}, \text{pa}, \text{ps}\}$Grow). The difference in our
measurements between pool growing (pxGrow) and the corresponding variants with
enslavement (uxGrow) are not very big. Growing with marking performs better than globally
synchronized growing except for update heavy workloads.  The price of
growing compared to a fixed size is less than a factor of two for insertions and
updates (aggregation) and negligible for find-operations.  Moreover, this
slowdown is comparable to the slowdown experienced in sequential hash tables
when growing is necessary.  None of the other data structures that support
growing comes even close to our data structures.  For insertions and updates we
are an order of magnitude faster then many of our competitors. Furthermore, only
one competitor achieves speedups above one when inserting into a growing table
(junction grampa).  

Among the tested hash tables, only TBB, Cuckoo, and RCU have the ability to
store arbitrary key-/value-type combinations.  Therefore, using arbitrary data
objects with one of these hash tables can be considered to cost at least an
order of magnitude in performance
($\text{TBB}[\textit{arbitrary}] \leq \text{TBB}[\textit{word sized}] \approx 1/10\cdot
\text{xyGrow}$).
In our opinion, this restricts the use of these data structures to situations
where hash table accesses are not a computational bottleneck.  For more
demanding applications the only way to go is to get rid of the general data
types or the need for concurrent hash tables altogether. We believe that the
generalizations we have outlined in \autoref{sec:add_string_keys} will be able
to close this gap. Actual implementations and experiments are therefore
interesting future work.

Finally let us consider the situation where we need general data types but no
growing. Again, all the competitors are an order of magnitude slower for
insertion than our bounded hash tables. The single exception is cuckoo, which is
only five times slower for insertion and six times slower for successful
reads. However, it severely suffers from contention being an almost record
breaking factor of 5\,600 slower under find-operations with contention.  Again,
it seems that better data structures should be possible.


\section{Conclusion}\label{sec:conclusion}

We demonstrate that a bounded linear probing hash table specialized to pairs of
machine words has much higher performance than currently available general
purpose hash tables like Intel TBB, Leahash, or RCU based implementations. This is
not surprising from a qualitative point of view given previous publications
\cite{StivalaCASTable10,kim2013performance,shun2014phase}. However, we found it
surprising how big the differences can be in particular in the presence of
contention. For example, the simple decision to require a lock for reading can
decrease performance by almost four orders of magnitude.

Perhaps our main contribution is to show that integrating an adaptive growing
mechanism into that data structure has only a moderate performance
penalty. Furthermore, the used migration algorithm can also be used to implement
deletions in a way that reclaims freed memory.  We also explain how to further
generalize the data structure to allowing more general data types.

The next logical steps are to implement these further generalizations
efficiently and to integrate them into an easy to use library that hides most of
the variants from the user, e.g., using programming techniques like partial
template specialization.

Further directions of research could be to look for a practical growable
lock-free hash table.  




\paragraph*{Acknowledgments}
We would like to thank Markus Armbruster, Ingo M\"uller, and Julian Shun for
fruitful discussions.

\bibliographystyle{plain}










\end{document}